\documentclass[twocolumn,prX,groupedaddress,showpacs,amsmath,amssymb]{revtex4}
\usepackage{float}
\usepackage{graphicx}

\begin{document}

\title{Quantum noise in large-scale coherent nonlinear photonic circuits}
\author{Charles Santori
\email{charles.santori@hp.com}}
\author{Jason S. Pelc}
\author{Raymond G. Beausoleil}
\affiliation{Hewlett-Packard Laboratories, 1501 Page Mill Road, MS1123, Palo Alto, California 94304, USA}
\author{Nikolas Tezak}
\author{Ryan Hamerly}
\author{Hideo Mabuchi}
\affiliation{Edward L. Ginzton Laboratory, Stanford University, Stanford, CA 94305, USA}

\newcommand{\wig}{\mathcal{W}}		
\newcommand{\bin}{\beta_{\rm in}}	
\newcommand{\bout}{\beta_{\rm out}}
\newcommand{\bfield}{\beta}
\newcommand{\binn}[1]{\beta_{{\rm in},{#1}}}	
\newcommand{\boutt}[1]{\beta_{{\rm out},{#1}}}
\newcommand{\bincoh}{\bar{\beta}_{\rm in}}
\newcommand{\bincohh}[1]{\bar{\beta}_{{\rm in},#1}}
\newcommand{\Lphase}{\psi}		
\newcommand{\z}{\alpha}			
\newcommand{\zs}{\z^\ast}			

\begin{abstract}

A semiclassical simulation approach is presented for studying quantum noise in large-scale photonic circuits incorporating an ideal Kerr nonlinearity. A circuit solver is used to generate matrices defining a set of stochastic differential equations, in which the resonator field variables represent random samplings of the Wigner quasi-probability distributions. Although the semiclassical approach involves making a large-photon-number approximation, tests on one- and two-resonator circuits indicate satisfactory agreement between the semiclassical and full-quantum simulation results in the parameter regime of interest. The semiclassical model is used to simulate random errors in a large-scale circuit that contains 88 resonators and hundreds of components in total, and functions as a 4-bit ripple counter. The error rate as a function of on-state photon number is examined, and it is observed that the quantum fluctuation amplitudes do not increase as signals propagate through the circuit, an important property for scalability.

\end{abstract}
\maketitle

\section{Introduction}

While all-optical logic has historically been difficult to implement~\cite{ref:miller2010aot}, recent progress in micro- and nano-scale optical devices has renewed interest in this subject, as it could allow pushing energy consumption to regimes that have not been attainable in bulk optical systems. In room-temperature experiments, carrier-based optical switching has been demonstrated at $100\,\mathrm{fJ}$ pulse energies in silicon devices~\cite{ref:tanabe2005aos} and at sub-fJ pulse energies in devices made from III-V materials~\cite{ref:nozaki2010sfa}, where bistable optical memories have also been implemented \cite{ref:nozaki2012upa}.  Switching based on far-off-resonant (Kerr) nonlinearities typically requires higher powers, but has been demonstrated at $< 1\,\mathrm{pJ}$ in ring resonators made from amorphous silicon~\cite{ref:pelc2013pao}. In low-temperature experiments, fJ-scale optical logic has been achieved in exciton-polariton systems~\cite{ref:ballarini2013aop}, and switching near the single-photon level has been achieved in quantum dots coupled to photonic-crystal resonators\cite{ref:faraon2008cgn,ref:bose2012lpn}. Low-temperature atomic ensembles can also be used quite effectively for low-photon-number switching~\cite{ref:chen2013aos,ref:kwon2013fsa}. Thus, low-temperature switching experiments are already well into regimes where quantum effects are important for the switching dynamics~\cite{ref:kerckhoff2011rsb}, and room-temperature devices may soon reach such regimes, as well.

Simulation tools are therefore needed that can predict quantum effects in the $\sim 10-1000$ photon regime, such as random error rates due to quantum jumps, and new types of behavior that might occur when many components interact through coherent signals. Full-quantum simulation methods, such as the SLH model (the letters referring to scattering, collapse and Hamiltonian operators)~\cite{ref:gough2009spa} and its implementation within the Quantum Hardware Description Language (QHDL)~\cite{ref:tezak2012spc} may be used to study circuits containing 1-3 idealized components, but the exponential scaling of the state space with respect to the number of resonators requires some kind of approximation to be made before larger circuits can be studied~\cite{ref:tezak2012spc,ref:bouten2008alt,ref:mabuchi2008dmb,ref:nielsen2009qfr}.

Here, we describe semiclassical simulations following a method that can be applied to large-scale networks of Kerr-nonlinear resonators connected through linear optics. The stochastic differential equations we use are similar to equations used previously~\cite{ref:carter1995qtn}, with the dynamic field variables representing a random sampling of the Wigner quasi-probability distribution. However, this is the first demonstration, to our knowledge, of combining such a model with a circuit solver that automatically converts a netlist, which describes an optical circuit topology, into a set of matrices representing the stochastic differential equations (see Section~\ref{sec:quantum_semiclassic_comparison} for a more detailed discussion of the netlist). This allows us to construct large circuits based on multiple layers of subcircuits, such as the 4-bit ripple counter described below. The computation time scales polynomially with the number of components, and in many circuits the scaling is approximately linear.  Even though our model uses just one complex variable to describe each resonator, for the parameter regime of interest, we can reproduce the spontaneous switching events in one- or two-element circuits predicted by a full-quantum simulation. When the noise terms are removed, our model is the same as coupled mode theory~\cite{ref:haus1991cmt} (with energy scaled in photon units), and thus the same tool can be used to perform classical simulations. Our results indicate that a pure Kerr nonlinearity, combined with auxiliary coherent inputs, can be used to create arbitrary logic, with tolerance to moderate linear loss, and adequate signal restoration for cascading.  Such circuits can function with negligible errors at intra-cavity energies down to a few tens of attojoules.

\section{Simulation Method}
\label{sec:simulation_methods}

Suppose we have an optical circuit consisting of a set of nonlinear resonators connected by linear optical components, including waveguides, beamsplitters and phase shifters, and driven by coherent external inputs. When simulating such a circuit we will be interested in characterizing both its internal state as well as the resulting output fields. In modeling the internal state we will only keep track of long-lived resonator modes within the photonic structure and assume that the fields propagating in the interconnecting waveguides can always be described in terms of these resonator modes and the input fields at the same instant in time. This approximation, made to simplify the computational treatment, can be used for circuits with short connections between components, so that the circuit dynamics occur on timescales much longer than the time it takes light to propagate across the structure. It does not allow simulation of circuits where the propagation delays are comparable to the resonator lifetimes (typically picoseconds), or where long delay lines have been intentionally introduced (in our treatment a delay line is just a phase shift). In the future, the ability to simulate these types of circuits could be implemented to first approximation (neglecting dispersion) by introducing delays in the differential equations.

Since we are interested in operating at fairly low light intensities, quantum shot noise can play an important role in the dynamics. Therefore our model is of a stochastic nature: the resonator fields as well as the inputs and outputs are taken to be stochastic processes. In Section \ref{sec:wigner_derivation} we will show how the stochastic differential equation (SDE) describing a resonator coupled to a single waveguide can be derived from a Fokker-Planck equation for the Wigner quasi-probability distribution. In Section \ref{sec:equations}, we extend this to multiple inputs and outputs, and present the SDE that is the basis for our semiclassical model. In \ref{staticcomponents} we give the input-output relations for some static components, and in \ref{sec:circuit_eqs} we show an algebraic approach for converting a circuit containing many components into a set of SDE coupling matrices. Finally, in \ref{sec:quantum_semiclassic_comparison} we discuss the software and numerical implementation of our model, and compare its result with those from full quantum simulations.

\subsection{Single-mode resonator coupled to one waveguide}
\label{sec:wigner_derivation}
We start with the simplest open-system Kerr-nonlinear resonator model featuring a single mode and a single dissipative coupling to an external field in the vacuum.
This system is described by a Hamiltonian $H$ and a single collapse operator $L$ given by
\begin{align}
  H & = \Delta a^\dagger a + \chi a^{\dagger 2}a^{2} \\
   L &= \sqrt{\kappa}e^{i\Lphase} a
\end{align}
where $a^\dagger$ and $a$ are photon creation and annihilation operators for the resonator, $L$ is the Lindblad-collapse operator~\cite{ref:gardiner2004qn} associated with photon leakage out of the resonator at a rate $\kappa$, $\Lphase$ is the coupling phase, $\Delta$ is the resonator detuning from a reference frequency, and $\chi$ is the nonlinearity.  The corresponding Lindblad-Master equation is given by
\begin{align}
\dot{\rho} = -i [H, \rho] + L \rho L^{\dagger} - \frac{1}{2}\left(L^{\dagger}L \rho + \rho L^\dagger L \right),
\end{align}
consisting of the usual coherent part $-i [H, \rho]$ and the dissipative part $\mathcal{D}[L]\rho :=  L \rho L^{\dagger} - \frac{1}{2}\left[L^{\dagger}L \rho + \rho L^\dagger L \right]$.

The well-known Wigner transform of an oscillator state $\rho$ is given by~\cite{ref:scully1997qo}
\begin{align}
\wig\{\rho\}(\z,\zs) &:= \frac{1}{\pi^{2}}\int d^{2}\beta e^{-i\beta\zs - i \beta^{\ast}\z}{\rm Tr}\left(e^{i\beta a^{\dagger}+i\beta^{\ast}a}\rho\right),
\end{align}
and from this we define the system's Wigner function as $W(\z, \z^\ast, t):=\wig\{\rho(t)\}$.
To find out how the Wigner function evolves in time, we differentiate and substitute in the Master equation
\begin{align*}
\partial_tW(\z, \zs,t) = \wig\left\{\frac{d}{dt}\rho\right\} = \wig\left\{ -i [H, \rho]\right \} + \wig\left\{\mathcal{D}[L]\rho\right\}.
\end{align*}
The Wigner distribution is useful because of the correspondence between quantum expectation values for mode operator moments and its moments, e.g.
\begin{align}
  \langle a \rangle & = \langle \z \rangle_W, \\
  \langle a^\dagger \rangle & = \langle \z^\ast \rangle_W, \\
  \langle n \rangle & = \langle a^\dagger a \rangle =  \langle \z^\ast \z \rangle_W - \frac{1}{2},  \\
  {\rm Var}(n) & = \langle \left(n - \langle n\rangle\right)^2\rangle = {\rm Var}(\z^\ast \z)_W -\frac{1}{4}\\
  {\rm Var}\left(\frac{a + a^\dagger}{2}\right) &= {\rm Var}\left(\frac{\z + \z^\ast}{2}\right)_W.
\end{align}
This means that instead of performing a full quantum simulation to
evaluate operator expectation values, we can instead sample directly
from the Wigner-distribution.

A few calculations~\cite{ref:carter1995qtn} show that
{\small\begin{align*}
	\wig\{a \rho \} &= \left(\z + \frac{\partial_{\zs}}{2}\right) \wig\{\rho\}, &
	\wig\{ \rho a \} &= \left(\z - \frac{\partial_{\zs}}{2}\right) \wig\{\rho\},\\
	\wig\{a^{\dagger}\rho \} &= \left(\zs - \frac{\partial_{\z}}{2}\right) \wig\{\rho\}, &
	\wig\{\rho a^{\dagger} \} &= \left(\zs + \frac{\partial_{\z}}{2}\right) \wig\{\rho\},
\end{align*}}
and these relations can be iterated to yield
\begin{align*}
    \wig\{[a^{\dagger}a, \rho]\} 
    & = \left(\partial_{\zs} \zs - \partial_{\z} \z\right) \wig\{\rho\}, \\
 \wig\{[a^{\dagger 2}a^{2}, \rho]\}
    & = 2\Huge[\partial_{\z^\ast}(\zs\z- 1)\z^\ast - \partial_{\z}(\z^\ast\z- 1)\z  \nonumber  \\
&\quad + \frac{1}{4}\partial_{\z}^{2}\partial_{\z^\ast}\z - \frac{1}{4}\partial_{\zs}^{2}\partial_{\z}\zs\Huge]\wig\{\rho\}, \\
\wig\{\mathcal{D}\left[a\right]\rho\} & = \frac{1}{2} \left[\partial_{\z^\ast}\z^\ast +  \partial_{\z}\z +\partial_{\z^\ast}\partial_{\z}\right]\wig\{\rho\},
\end{align*}
such that we ultimately find
\begin{align} \nonumber
  \partial_t\wig &= -i\Delta\left(\partial_{\zs} \zs - \partial_{\z} \z\right) \wig \\ \nonumber
& \quad -2i\chi \left(\partial_{\z^\ast}(\zs\z- 1)\z^\ast - \partial_{\z}(\z^\ast\z- 1)\z\right)\wig   \\ \nonumber
& \quad -2i\chi\left( \frac{1}{4}\partial_{\z}^{2}\partial_{\z^\ast}\z - \frac{1}{4}\partial_{\zs}^{2}\partial_{\z}\zs\right)\wig \\ 
& \quad +\frac{\kappa}{2} \left(\partial_\z \z + \partial_{\z^\ast} \z^\ast + \partial_{\z^\ast}\partial_{\z}\right)\wig. \label{eq:singlemodewignerderiv}
\end{align}
As discussed in Ref.~\cite{ref:carter1995qtn}, in order to arrive at proper Fokker-Planck equations, we must drop the third-order derivatives, i.e. the third row in Equation~\eqref{eq:singlemodewignerderiv}. This is justified in the case of large photon numbers in the resonators, since, assuming $\partial_\z \sim 1$, the first-derivative terms containing $\chi$ are a factor of $\sim |\alpha|^2$ larger.  The remaining first and second-derivative terms can be considered as representing drift and diffusion, respectively, in a stochastic process.  We can rewrite the first and second-derivative terms of Eq.~\ref{eq:singlemodewignerderiv} as,
\begin{equation}
\partial_t \wig \approx -\sum_p \partial_{\z_p} \left( A_p W \right) + \frac{1}{2} \sum_{p,p^\prime} \partial_{\z_p}\partial_{{\z}_{p^\prime}} \left[ {\left(B B^T \right)}_{p,p^\prime} W \right]
\label{eq:fokkerplanckfinal}
\end{equation}
where ${\z}_p$, with $p = \{r, i\}$, denotes the real or imaginary part of $\alpha$, and
\begin{eqnarray}
A_r &=& -\frac{\kappa}{2} \alpha_r + \left(\Delta + 2\chi(\alpha_r^2 + \alpha_i^2 - 1)\right)\alpha_i \\
A_i &=& -\frac{\kappa}{2} \alpha_i - \left(\Delta + 2\chi(\alpha_r^2 + \alpha_i^2 - 1)\right)\alpha_r \\
B_{p,p^\prime} &=& \delta_{p, p^\prime} \frac{\sqrt{\kappa}}{2}
\end{eqnarray}
The stochastic equation corresponding to Eq.~\ref{eq:fokkerplanckfinal} (see Appendix B of Ref.~\cite{ref:carter1995qtn}) is,
\begin{equation}
d{\z}_p = A_p dt + \sum_{p^\prime} B_{p,p^\prime} dW_{p^\prime}
\end{equation}
where the noise increments $dW_r$ and $dW_i$ are taken as independent, zero-mean, Gaussian noise processes with $\langle dW_p(t_1)dW_{p^\prime}(t_2)\rangle = \delta_{p,p^\prime} \delta_{t_1,t_2} dt$. Combining the above expressions, the Langevin equation for $\alpha = \alpha_r + i \alpha_i$ is,
\begin{align}
\label{eq:singlemodesde}
  \dot{\z}(t) & = -\left[\frac{\kappa}{2} + i \Delta  + 2i \chi \left(\z^\ast(t)\z(t)-1\right)\right]\z(t) \\ \nonumber
& \quad - \sqrt{\kappa} e^{-i\Lphase} \bin(t)
\end{align}
where $\bin(t)$ is a complex Wiener process $\bin(t) = \eta^{(1)}(t) + i\eta^{(2)}(t)$, with $\langle\eta^{(m)}(t)\eta^{(n)}(t')\rangle = \frac{1}{4}\delta_{mn}\delta(t-t')$. Note that we have inserted a phase factor of $-e^{-i\Lphase}$, which has no effect on the stochastic process. However, with this phase factor, driving the resonator with an arbitrary coherent field (rather than using a vacuum state as we have done so far) displaces $\bin(t) \to \bincoh(t) + \eta(t)$, where $\bincoh(t)$ is a complex-valued, deterministic function of time, equal to the input field amplitude. We interpret $\bin(t)$ as representing the input field in the waveguide, which includes the quantum noise of a coherent state.
  
To obtain the output field in the waveguide, we can use the input-output formalism~\cite{ref:gardiner1985iod,ref:gardiner1992wfq}, which describes how input and output fields are related to each other and to a scattering system. For a single waveguide coupled to a system via a Heisenberg-picture coupling operator $L(t)$, the input-output relation is given as $b_{\rm out}(t) = b_{\rm in}(t) + L(t)$, where $b_{\rm in}(t)$ and $b_{\rm out}(t)$ are quantum operators representing the input and ouput fields in the waveguide. For our system this leads to,
\begin{align}\label{eq:singlemodeout}
  \bout(t) = \sqrt{\kappa} e^{i\Lphase} \z(t) + \bin(t).
\end{align}

Alternatively, Eqs.~\ref{eq:singlemodesde}-\ref{eq:singlemodeout} can be derived starting from the Hamiltonian of a nonlinear resonator coupled to a continuum of waveguide modes. In this picture there are no collapse operators. One first derives a set of Fokker-Planck equations for both the resonator and waveguide modes. It is again necessary to drop the third-derivative terms associated with the Kerr nonlinearity. However, there are no second-derivative terms~\cite{ref:carter1995qtn}. The input and output fields $b_{\rm in}(t)$ and $b_{\rm out}(t)$ are then defined as Fourier sums of the waveguide modes before and after interaction with the resonator.  Since the external modes are treated from the beginning as quantum objects, the noise in $\bin(t)$ enters directly through the quantum states of the external inputs. In this derivation the inputs need not be coherent states, though highly nonclassical states with negative Wigner functions are still not allowed.

\subsection{Single-mode resonators with multiple inputs and outputs}
\label{sec:equations}
With the above work, the generalization to multiple inputs and outputs is straightforward. We thus assume a Hamiltonian and vector of collapse operators
\begin{align}
  H & = \Delta a^\dagger a + \chi a^{\dagger 2}a^{2}
\\
  \mathbf{L} &= 
\begin{pmatrix}
\sqrt{\kappa_{1}}e^{i\Lphase_1} a \\ 
\sqrt{\kappa_{2}}e^{i\Lphase_2} a \\ 
\vdots \\
\sqrt{\kappa_{n}}e^{i\Lphase_n} a 
\end{pmatrix}
\end{align}
where we have allowed for different phases associated with each input port.
Inserting these into the Lindblad master equation:
\begin{align}
\dot{\rho} = -i [H, \rho] + \sum_{k=1}^n \mathcal{D}[L_{k}]\rho,
\end{align}
leads to the following SDE:
\begin{align}
\label{eq:eomsinternal}
  \dot{\z}(t) & = -\left[
\frac{\kappa}{2} + i \Delta + 2i\chi  (\zs \z - 1)\right] \z \\ \nonumber
& \quad - \sum_j \sqrt{\kappa_j}e^{-i\Lphase_j}\binn{j}(t) 
\\ \label{eq:eomsoutput1}
\boutt{j}(t) & =
   \sqrt{\kappa_j}e^{i\Lphase_j}\z(t) + \binn{j}(t)
\end{align}
where $\kappa = \sum_j \kappa_k$. Eqs.~\ref{eq:eomsinternal}-\ref{eq:eomsoutput1} are the starting point for our semiclassical simulations. The multiple inputs and outputs can represent either waveguides or free-space modes, as needed to describe scattering loss, for example. If the input field $\binn{j}(t)$ originates from outside the circuit, we set $\binn{j}(t) = \bincohh{j}(t) + \eta_j(t)$ as discussed above, where $\bincohh{j}(t)$ is a deterministic coherent amplitude and $\eta_j(t) = \eta_{j}^{(1)}(t) + i \eta_{j}^{(2)}$ is a complex gaussian noise process with zero mean $\langle \eta_j(t)\rangle = 0$ and second-order moments $\langle \eta_{j}^{(m)}(t)\eta_{j}^{(n)}(t')\rangle =\frac{1}{4}\delta_{jk}\delta_{mn}\delta(t-t')$. Alternatively, the input of one resonator may be supplied by the output of another resonator. The resulting coupled equations of motion are a system Langevin equations, i.e. stochastic differential equations, but since the coefficients to the noise terms are state-independent, they assume the same form in both the Ito and Stratonovich convention.

It is important to note that although the internal mode variables have
nonlinear equations of motion, the coupling to the external inputs
and noises is fully linear. This makes it straightforward to derive rules for how
to combine such systems into a circuit.

\subsection{Static Components}
\label{staticcomponents}

Besides resonators, three static components are needed -- a beamsplitter, a phase shifter and a coherent displacement.  These components do not have internal states; their input-output relations are fully described by a scattering matrix.  For the beamsplitter, which is parameterized by an angle $\theta$:
\begin{align}
	\left[\begin{array}{c} \boutt{1}(t) \\ \boutt{2}(t)\end{array}\right] &= 
		\left[\begin{array}{cc} \cos\theta & -\sin\theta \\ \sin\theta & \cos\theta \end{array}\right]
		\left[\begin{array}{c} \binn{1}(t) \\ \binn{2}(t)\end{array}\right]
\end{align}
A phase shifter is parameterized by a phase $\phi$:
\begin{align}
	\bout &= e^{i\phi}\bin
\end{align}
A coherent displacement is parameterized by a displacement field $\beta_0$, which adds to the current field.  This is equivalent to bouncing light off a highly reflective beamsplitter with a very strong field entering through the dark port:
\begin{align}
	\bout &= \bin + \beta_0
\end{align}

\subsection{Circuits of Components}
\label{sec:circuit_eqs}
Stochastic equations for circuits of many components can be obtained in a straightforward, algorithmic manner.  First, the input-output equations for each component $\mathcal{K}^{(i)}$ are written down in the following general form:
\begin{align}
\dot{\z}^{(i)}(t) &= \left[\mathbf{A}^{(i)} \z^{(i)}(t) + \mathbf{a}^{(i)} + A_{\rm NL}^{(i)}(\z^{(i)},t)\right]
	 + \mathbf{B}^{(i)} \bin^{(i)}(t) \nonumber \\
	\bout^{(i)}(t) &= \left[\mathbf{C}^{(i)} \z^{(i)}(t) + \mathbf{c}^{(i)}\right] + \mathbf{D}^{(i)}\bin^{(i)}(t) \label{eq:abcd-io}
\end{align}
where $i$ is the component index, $\z^{(i)}(t)$ is a vector of field amplitudes for all resonators belonging to $\mathcal{K}^{(i)}$; $\bin^{(i)}(t)$ and $\bout^{(i)}(t)$ are vectors of inputs and outputs to $\mathcal{K}^{(i)}$, $\mathbf{A}, \mathbf{B}, \mathbf{C}$, and $\mathbf{D}$ are constant matrices, and $\mathbf{a}$ and $\mathbf{c}$ are constant vectors. The vector $A_{\rm NL}^{(i)}$ describes the resonator nonlinearities.  For static components, only $\mathbf{D}$ and $\mathbf{c}$ are defined, the rest of the matrices and vectors being ignored because the component has no internal state.

A circuit consists of many such components $(\mathcal{K}^{(1)},\ldots,\mathcal{K}^{(n)})$, connected together, meaning that $\beta_{\mathrm{in},m}^{(i)}(t) = \beta_{\mathrm{out},n}^{(j)}(t)$ for particular values of $i$, $j$, $m$ and $n$. A simple concatenation, in which the components connect only to external fields, obeys equations of motion of the same form (\ref{eq:abcd-io}), with the following matrices:
{\small\begin{align}
	\z &= \left[\begin{array}{c}\!\z^{(1)}\! \\ \vdots \\ \!\z^{(n)}\!\end{array}\right],\ \
	A_{\rm NL}(z,t) = \left[\begin{array}{c}\!A_{\rm NL}^{(1)}(\z^{(1)},t)\! \\ \vdots \\ \!A_{\rm NL}^{(n)}(\z^{(n)},t)\!\end{array}\right],\nonumber \\	 
	\mathbf{A} &= \left[\begin{array}{ccc}\mathbf{A^{(1)}} & \!0\! & 0 \\ 0 & \!\!\ddots\!\! & 0 \\ 0 & \!0\! & \mathbf{A^{(n)}}\end{array}\right],\ \ 
	\mathbf{a} = \left[\begin{array}{c}\!\mathbf{a^{(1)}}\! \\ \vdots \\ \!\mathbf{a^{(n)}}\!\end{array}\right],\ \
	\mathbf{B} = \left[\begin{array}{ccc}\mathbf{B^{(1)}} & \!0\! & 0 \\ 0 & \!\!\ddots\!\! & 0 \\ 0 & \!0\! & \mathbf{B^{(n)}}\end{array}\right],\nonumber \\ 
	\mathbf{C} &= \left[\begin{array}{ccc}\mathbf{C^{(1)}} & \!0\! & 0 \\ 0 & \!\!\ddots\!\! & 0 \\ 0 & \!0\! & \mathbf{C^{(n)}}\end{array}\right],\ \ 
	\mathbf{c} = \left[\begin{array}{c}\!\mathbf{c^{(1)}}\! \\ \vdots \\ \!\mathbf{c^{(n)}}\!\end{array}\right],\ \
	\mathbf{D} = \left[\begin{array}{ccc}\mathbf{D^{(1)}} & \!0\! & 0 \\ 0 & \!\!\ddots\!\! & 0 \\ 0 & \!0\! & \mathbf{D^{(n)}}\end{array}\right]
\end{align}}

\begin{figure}[t]
\centering
\includegraphics[width=3.3in]{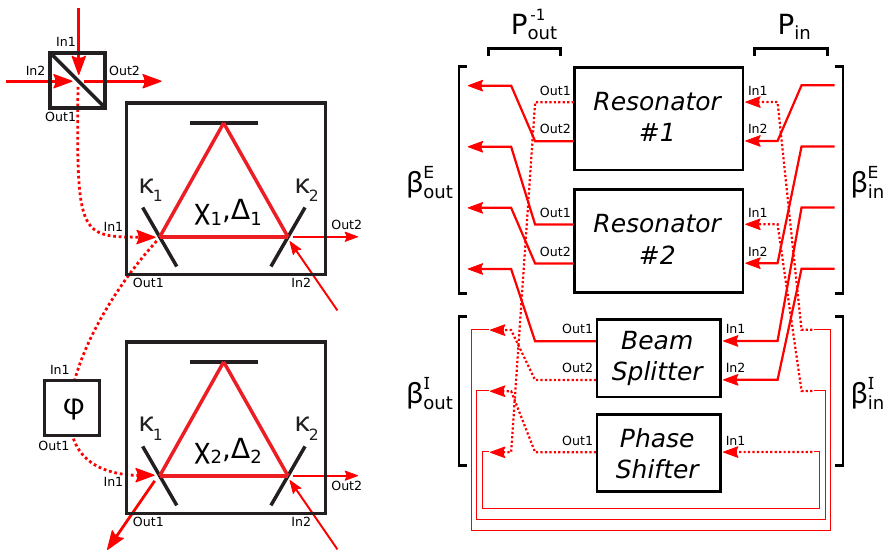}
\caption{Example circuit showing external ($\bout^E,\bin^E$) and internal ($\bout^I, \bin^I$) input-output fields.}
\label{fig:qhdl_fig1}
\end{figure}

Connections between components are modeled by splitting the I/O fields into ``internal'' ($\bfield^{I}$) fields and ``external'' ($\bfield^E$) fields, as follows:
\begin{equation}
	\bin = P_{\rm in} \left[\begin{array}{c} \bin^E \\ \bin^I \end{array}\right],\ \ \ \ \bout = P_{\rm out} \left[\begin{array}{c} \bout^E \\ \bout^I \end{array}\right]
\end{equation}
where $P_{\rm in}$ and $P_{\rm out}$ are permutation matrices.  This is illustrated with the toy circuit in Figure \ref{fig:qhdl_fig1}.  In this circuit, there are four components, four external fields (solid red lines) and three internal fields (dotted lines).  These permutations allow $\mathbf{B}$, $\mathbf{C}$ and $\mathbf{D}$ to be written in block form that separates their ``internal'' behavior from their ``external'' behavior:
\begin{align}
    \mathbf{B} P_{\rm in} &= \left[\begin{array}{cc} \mathbf{B}^E & \mathbf{B}^I\end{array}\right] \nonumber \\
    P_{\rm out}^T \mathbf{C} &= \left[\begin{array}{c}\mathbf{C}^E \\ \mathbf{C}^I\end{array}\right] \nonumber \\
    P_{\rm out}^T \mathbf{c} &= \left[\begin{array}{c}\mathbf{c}^E \\ \mathbf{c}^I\end{array}\right] \nonumber \\
    P_{\rm out}^T \mathbf{D} P_{\rm in} &= \left[\begin{array}{cc}\mathbf{D}^{EE} & \mathbf{D}^{EI} \\ \mathbf{D}^{IE} & \mathbf{D}^{II} \end{array}\right]
\end{align}
The internal fields are arranged so that $\boutt{k}^I$ connects to $\binn{k}^I$.  Making these connections is equivalent to imposing $\bin^I = \bout^I$.  One can then solve for the internal fields and eliminate them.  The equations of motion for the circuit become:
\begin{align}
    \dot{\z} &= \left[\mathbf{A} + \mathbf{B}^I(1 - \mathbf{D}^{II})^{-1}\mathbf{C}^I\right] \z \nonumber \\
    & \quad + \left[\mathbf{a} + \mathbf{B}^I(1 - \mathbf{D}^{II})^{-1}\mathbf{c}^I\right] \nonumber + A_{\rm NL}(\z)\\
    & \quad + \left[\mathbf{B}^E + \mathbf{B}^I(1 - \mathbf{D}^{II})^{-1}\mathbf{D}^{IE}\right]\bin^E \\
	\bout^E & = \left[\mathbf{C}^E + \mathbf{D}^{EI}(1 - \mathbf{D}^{II})^{-1}\mathbf{C}^I\right] \z \nonumber \\
	& \quad + \left[\mathbf{c}^E + \mathbf{D}^{EI}(1 - \mathbf{D}^{II})^{-1}\mathbf{c}^I\right] \nonumber \\
	& \quad + \left[\mathbf{D}^{EE} + \mathbf{D}^{EI}(1 - \mathbf{D}^{II})^{-1}\mathbf{D}^{IE}\right]\bin^E
\end{align}
The nonlinear part of the stochastic equations, $A_{\rm NL}(\z)$, does not change.  In other words, the only effect of interconnections is to renormalize the linear part of the input-output equations, as follows:
\begin{align}
	\mathbf{A} &\rightarrow \left[\mathbf{A} + \mathbf{B}^I(1 - \mathbf{D}^{II})^{-1}\mathbf{C}^I\right] \nonumber \\
	\mathbf{a} &\rightarrow \left[\mathbf{a} + \mathbf{B}^I(1 - \mathbf{D}^{II})^{-1}\mathbf{c}^I\right] \nonumber \\
	\mathbf{B} &\rightarrow \left[\mathbf{B}^E + \mathbf{B}^I(1 - \mathbf{D}^{II})^{-1}\mathbf{D}^{IE}\right] \nonumber \\
	\mathbf{C} &\rightarrow \left[\mathbf{C}^E + \mathbf{D}^{EI}(1 - \mathbf{D}^{II})^{-1}\mathbf{C}^I\right] \nonumber \\
	\mathbf{c} &\rightarrow \left[\mathbf{c}^E + \mathbf{D}^{EI}(1 - \mathbf{D}^{II})^{-1}\mathbf{c}^I\right] \nonumber \\
	\mathbf{D} &\rightarrow \left[\mathbf{D}^{EE} + \mathbf{D}^{EI}(1 - \mathbf{D}^{II})^{-1}\mathbf{D}^{IE}\right]
\end{align}

An alternative approach to generating the $\mathbf{A}, \mathbf{B}, \mathbf{C}, \mathbf{D}$ coupling matrices is to propagate backwards from a given component, accumulating amplitudes from other components, splitting into additional paths when needed, and terminating at external inputs. This method works well for circuits without loops, but may converge slowly if low-loss loops (effectively cavities without internal state) are present. The algebraic approach presented above is advantageous for such loop-containing circuits, such as those in Ref.~\cite{ref:huybrechts2010sbn}. Finally, it should be noted that instead of applying these circuit reduction rules, one can also work in a fully quantum picture and describe the whole network using SLH models~\cite{ref:gough2009spa}. In this case, the final overall network model leads to a master equation that leads to the same SDEs as above, although at greater computational cost, because the SLH formalism is more general and requires working explicitly with matrices whose elements are themselves non-commutative operators.

For circuits with fixed component parameters, the coupling  matrices need to be computed only once at the beginning of the simulation.  The computational difficulty of integrating the SDEs scales, at worst, quadratically with the number of resonators. However, for many circuits, the coupling matrices are sparse, and the scaling is expected to be nearly linear.

\subsection{Implementation and Validity}
\label{sec:quantum_semiclassic_comparison}
In the remainder of this paper we shall present simulation results obtained using a model based on Eqs.~\ref{eq:eomsinternal}-\ref{eq:eomsoutput1} with $\psi_i = -\pi/2$ (this determines the rotation angle of the Wigner distribution plots). The program, implemented in Matlab, allows a circuit to be defined as a netlist, which is a list of components, their parameters and connections. In the initial implementation, the allowed components are resonators (arbitrary number of input/ouput ports), two-port beamsplitters, phase shifters, non-operation (identity) components, external inputs, outputs, and custom-named compound components.  The compound components are similarly defined by netlists. Unless a component is an external input, a netlist entry must also specify the source of each of the component's inputs. Functions were written to flatten the netlist (by expanding compound components and connecting their inputs and outputs to the external circuit), compute circuit statistics and check for bad connections. We initially used a back-propagation method to convert the netlist to a set of matrices $\mathbf{A}, \mathbf{B}, \mathbf{C}, \mathbf{D}$ defining the stochastic differential equations. More recently, the algebraic approach described above has been implemented, and the two methods have been shown to produce consistent results. Integration of the stochastic equations was performed using an Euler-Maruyama timestep, modified to use exponential terms for the internal resonator dynamics:
\begin{equation}
\begin{split}
\alpha_{j}[n+1] = & \alpha_{j}[n] e^{(-i\Delta_j - \kappa_j/2 - 2i\chi_j |\alpha_{j}[n]|^2) \delta t} \\
 & + \left(\sum_k A_{jk} \alpha_{k}[n] + \sum_k B_{jk} \beta_{\mathrm{in},k}[n] \right) \delta t
\end{split}
\label{eq:timestep}
\end{equation}
where $\beta_{\mathrm{in},k}[n]$ includes a deterministic time-varying drive field plus independent Gaussian random variables for each timestep with amplitude $\sigma = 1/(2\sqrt{\delta t})$ for both the real and imaginary components, obtained from \verb|normrnd|.  The timestep $\delta t$ was set small enough that the spontaneous jump rates in single-resonator calculations appeared to be independent of $\delta t$. Typically $\delta t = 0.025 / \max(\{\Delta_j, \kappa_j \} )$ which equals $5 \times 10^{-4}$ for the counter circuit discussed below.

\begin{figure*}[ht]
\centering
\includegraphics[width=7in]{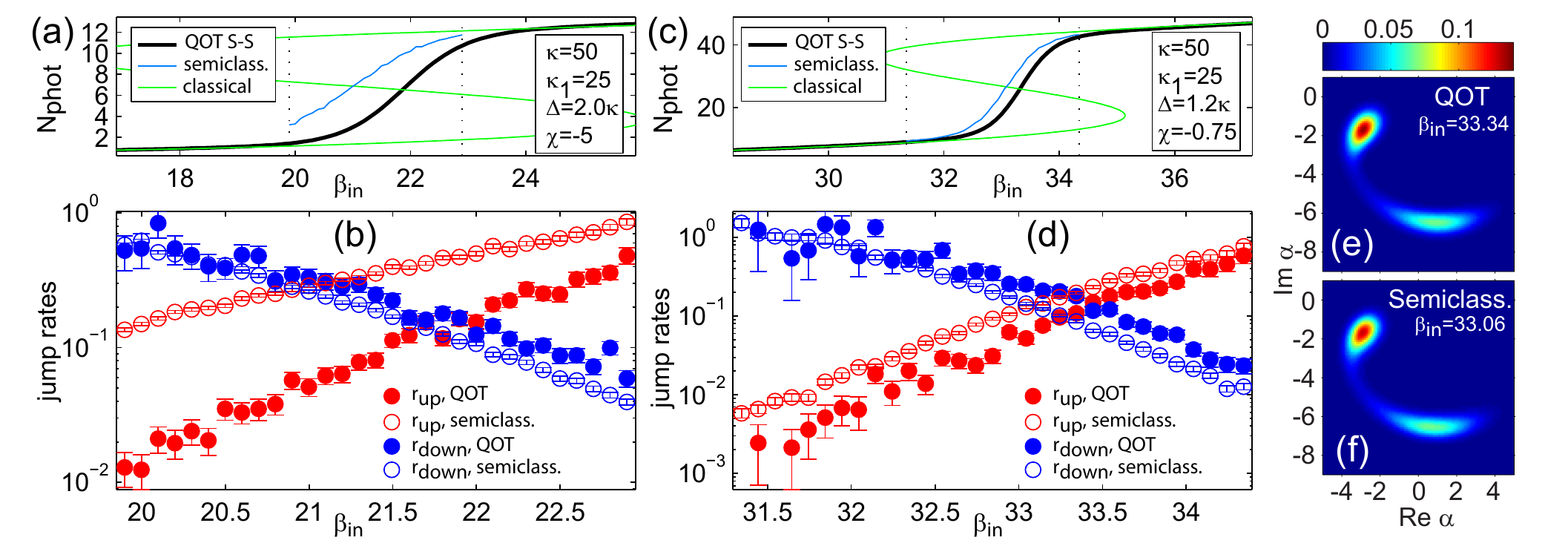}
\caption{(a) Time-averaged photon number vs. input field $\beta_\mathrm{in}$ as computed using the Quantum Optics Toolbox (QOT)~\cite{ref:tan1999ctq} steady-state solution, our semiclassical approximation, and the classical formula, in a low-photon-number regime (parameters given in figure). (b) Upward (red) and downward (blue) jump rates estimated from QOT Monte-Carlo integration (solid) and semiclassical (empty circles) simulations.  (c,d) Similar to (a,b), but in a higher photon-number regime, giving better agreement.  (e,f) Time-averaged Wigner distributions from the QOT and semiclassical simulations, with $\beta_\mathrm{in}$ set to obtain equal upper and lower state occupations; all other parameters same as in (c,d)}
\label{fig:qot_compare}
\end{figure*}
To check the validity of our semiclassical equations, we compared results with those obtained from a full quantum simulation, using the Quantum Optics Toolbox (QOT)~\cite{ref:tan1999ctq}. One type of comparison is to examine the spontaneous jump rates for a single, bistable Kerr resonator with a constant drive field.
 As discussed in more detail in Section~\ref{sec:amplifiers} below, when the drive frequency is sufficiently far detuned from the resonator, for a certain range of drive powers, the resonator can be in either of two states. In the lower-energy state, little light enters because of the large detuning. In the high-energy state, enough light has entered that the Kerr nonlinearity keeps the effective resonator frequency close to that of the input. Classically, these two states are both stable, but in the quantum regime, spontaneous jumps occur between them.

Fig.~\ref{fig:qot_compare} shows example results for a two-port resonator with $\kappa_1 = \kappa_2 = \kappa/2$, driven through one input. Fig~\ref{fig:qot_compare}a,b shows an example with a very strong nonlinearity, so that bistability occurs at very low photon numbers.  Fig~\ref{fig:qot_compare}a shows the time-averaged behavior, calculated using the steady-state density matrix solution from the QOT (black), by time-averaging the solution from our semiclassical model (blue), or using the classical formula in Eq.~\ref{eq:class_bistable} (green).  Fig~\ref{fig:qot_compare}b shows the jump rates, estimated by counting upward and downward transitions using our semiclassical model and using QOT's Monte-Carlo integration. The simulation length was $t_\mathrm{max} = 1000$ time units in the QOT calculation, and $t_\mathrm{max} = 10000$ in the semiclassical calculation. In this low-photon-number example, we see fairly good agreement in the downward jump rates $r_\mathrm{down}$, but the upward jump rates $r_\mathrm{up}$ are quite different. This leads also to a large difference in the time-averaged photon numbers. This disagreement is not surprising given that, in its lower state, the resonator contains just $\sim 2$ photons, and thus the approximation made in dropping the third-derivative terms from the Fokker-Planck equations (see Sec.~\ref{sec:wigner_derivation}) is not expected to be valid. In a somewhat higher-photon-number case (Fig~\ref{fig:qot_compare}c,d), the downward and upward rates both show good, though imperfect, agreement. The semiclassical results appear to be shifted horizontally from the QOT results by $\delta \beta_\mathrm{in} \sim 0.3$.  For a third comparison in which the ``on'' state had $\sim 95$ photons (using $\Delta=1.1\kappa$, $\chi=-0.3$), the jump-rate curves were again horizontally shifted, by $\delta \beta_\mathrm{in} \sim 0.35$~\footnote{For these higher photon numbers, the Quantum Optics Toolbox function mcsolve had to be recompiled for double-precision arithmetic to give accurate results.}.  The time-averaged Wigner functions computed from the full-quantum and semiclassical simulations (Fig.\ref{fig:qot_compare}e,f) look quite similar, provided that $\beta_\mathrm{in}$ is set to give the same ratio of upper and lower state populations in each case.

We also compared spontaneous jump rates in the two-resonator latch circuit presented below (Sec.~\ref{sec:latch}) for the QOT and semiclassical methods. For the QOT simulation we used the Hamiltonian and collapse operators given in the Supp. Info. of Ref.~\cite{ref:mabuchi2011nia}.  The results, included in Fig.~\ref{fig:latch_errors}, show good agreement between the full-quantum and semiclassical methods.

Although our implementation of the semiclassical model is not optimized for speed, even for a single resonator we see a large speed improvement compared with a full-quantum Monte-Carlo simulation. This is because the semiclassical model describes the resonator with a single complex number, while the full-quantum simulation requires a number of Fock-state amplitudes that increases with increasing expected photon number. For example, for the parameters used in Fig.~\ref{fig:qot_compare}d, simulating to $t_\mathrm{max}=1000$ requires approximately $11 \, \mathrm{s}$ using a laptop computer with an Intel i7 processor.  The corresponding QOT simulation time varies from $120\,\mathrm{s}$ to $540\,\mathrm{s}$ over the range of input field amplitudes shown in the plot. For the two-resonator latch circuit, simulating to $t_\mathrm{max}=1000$ requires approximately $50\,\mathrm{s}$ using the semiclassical model, vs. $14000\,\mathrm{s}$ using the QOT.  For the 88-resonator counter circuit described below, simulating to $t_\mathrm{max}=160$ using the semiclassical model requires approximately $180\,\mathrm{s}$ of computation time.

\section{Examples}

In this section we apply the simulation method described above to a set of optical circuits that could be of interest for switching and logic.  In \ref{sec:amplifiers} we investigate whether circuits composed only of beamsplitters, phase shifters, and resonators with an ideal Kerr nonlinearity can provide the gain and digital signal restoration needed for cascading. In \ref{sec:AND}-\ref{sec:Dflipflop} we introduce a set of building blocks that can be used for general-purpose combinatorial and sequential logic, and as an example application, in \ref{sec:counter} we simulate a 4-bit ripple counter containing 88 resonators.

\subsection{Inverting and non-inverting amplifiers}
\label{sec:amplifiers}
\begin{figure}[ht]
\centering
\includegraphics[width=3.3in]{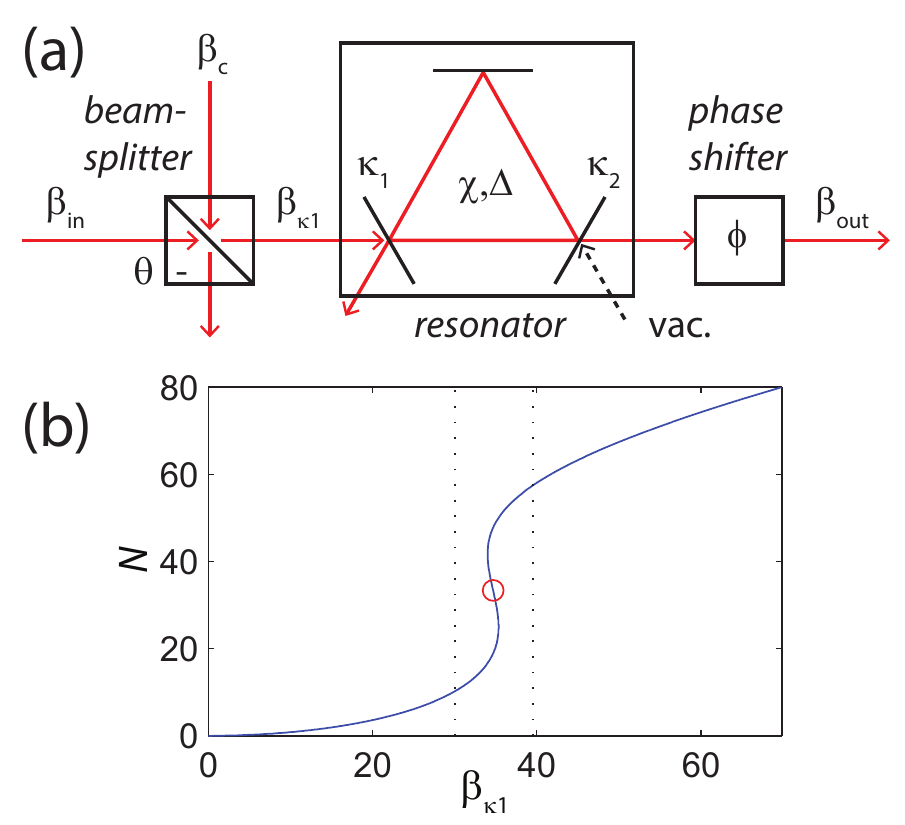}
\caption{(a) A simple amplifier circuit based on a nonlinear Kerr resonator. (b) Blue curve: Resonator photon number vs. resonator input field in the classical approximation, using the parameters from Table~\ref{tab:amplifiers}. Dotted line: designed ``low'' and ``high'' input fields. Red circle: inflection point.}
\label{fig:amplifier_circuit}
\end{figure}

Consider the amplifier circuit shown in Fig.~\ref{fig:amplifier_circuit}a. This is similar to the circuit in Ref.~\cite{ref:mabuchi2011nia}, but is simpler since it uses the signal from only one resonator output.  The input field $\beta_\mathrm{in}$ first interferes with a constant field $\beta_c$ on a beamsplitter, which has amplitude transmission coefficient $\cos \theta$ and reflection coefficient $\sin \theta$ (the minus sign in the lower-left corner of the beamsplitter indicates which output has a minus sign for the reflected component). The output exiting to the right is $\beta_{\kappa 1} = \beta_\mathrm{in} \cos \theta + \beta_c \sin \theta$.  This field then enters the first input of a two-port ring resonator. In this example we have not yet included intrinsic losses, so the total loss is the sum of the two coupling losses, $\kappa = \kappa_1 + \kappa_2$.  The second resonator output passes through a phase shifter with phase $\phi$ that is chosen so that the output has a phase of approximately zero in the ``high'' state.  This last component captures the necessity of controlling the phases between components. In an experimental realization, the phase shifter could correspond simply to a carefully chosen propagation length, or it could represent a tunable component. Whether or not active phase control is required depends on factors such as how reproducibly devices can be fabricated, and the amplitude of temperature fluctuations expected during circuit operation.

This circuit can function as either a non-inverting amplifier or an inverting amplifier, depending on the sign of the auxiliary field $\beta_c$.  Example parameters for both cases are given in Table~\ref{tab:amplifiers}.  Suppose, first, that we work in non-inverting mode, and that the design input amplitudes to stage 0 are $\beta_\mathrm{in} = \{0, 10\}$.  Since the beamsplitter is highly transmissive, it serves mainly to displace the field with little loss in amplitude, so that the field incident onto the resonator is $\beta_{\kappa 1} = \{30.04, 39.53\}$. The resonator then serves as a nonlinear filter, as represented by the blue curve in Fig.~\ref{fig:amplifier_circuit}b, which is the steady-state solution to Eqs.~\ref{eq:eomsinternal}-\ref{eq:eomsoutput1} without the noise terms:
\begin{equation}
\kappa_1 |\beta_\mathrm{in,1}|^2 \approx n \left[ (\kappa/2)^2 + (\Delta + 2 \chi n)^2 \right]
\label{eq:class_bistable}
\end{equation}
where $\beta_\mathrm{in,1}$ is the external field incident on one of the resonator inputs, $n \approx |\alpha|^2$ is the resonator photon number, and we have taken the limit $n \gg 1$. From this equation, one can show that classical bistability occurs for $\Delta > \frac{\sqrt{3}}{2} \kappa$, if $\chi<0$. The red circle marks an inflection point in the RHS of the above equation, $n_\mathrm{inflection} = -\frac{\Delta}{3\chi}$.  The parameter values used in Fig.~\ref{fig:amplifier_circuit}b are given in Table~\ref{tab:amplifiers} (stage 0).

With these parameters, the nonlinear resonator is near the onset of bistability and exhibits a threshold-like response.  The dotted lines in Fig.~\ref{fig:amplifier_circuit}b represent the designed resonator field amplitudes for ``low'' and ``high'' inputs.  The field exiting the resonator on the right has amplitude $\approx \sqrt{\kappa_2 n} \approx \{17, 37\}$ for low and high inputs, respectively, and thus the input field swing of $10$ has been amplified by a factor of 2. A non-ideal feature of this simple circuit is that the low output is nonzero and has a different phase than that of the high output.  The residual low output may cause complications when cascading components but can be eliminated by adding an interference path~\cite{ref:mabuchi2011nia}, as is included in the logic components introduced starting in \ref{sec:AND}.

\begin{table}[h]
\centering
\begin{tabular}{|c|c|c|c|c|}
\hline
\multicolumn{5}{|c|}{Amplifier Parameters} \\
\hline
stage & 0 & 1 & 2 & 3 \\
\hline
$t = \cos \theta$ & $\sqrt{0.9}$ & $\sqrt{0.9}$ & $\sqrt{0.9}$ & $\sqrt{0.9}$ \\  
$r = \sin \theta$ & $\sqrt{0.1}$ & $\sqrt{0.1}$ & $\sqrt{0.1}$ & $\sqrt{0.1}$ \\
$\chi$ & -0.5 & -0.5 & -0.5 & -0.5 \\
$\kappa_1$ & 25 & 50 & 100 & 200 \\
$\kappa_2$ & 25 & 50 & 100 & 200 \\
$\Delta$ & 50 & 100 & 200 & 400 \\
\hline
\multicolumn{5}{|c|}{Non-inverting} \\
\hline
design $\beta_\mathrm{in}$ & $\{0, 10\}$ & $\{17, 37\}$ & $\{26, 77\}$ & $\{45, 145\}$ \\
$\beta_c$ & 95 & 140 & 285 & 580 \\
$\phi$ & -3.42 & -3.42 & -3.42 & -3.42 \\
\hline
\multicolumn{5}{|c|}{Inverting} \\
\hline
design $\beta_\mathrm{in}$ & $\{0, 10\}$ & $\{17, 37\}$ & $\{32, 80\}$ & $\{62, 163\}$ \\
$\beta_c$ & -125 & -300 & -607 & -1215 \\
$\phi$ & -0.2 & -0.65 & -0.74 & -0.74 \\ 
\hline
\end{tabular}
\caption{Simulation parameters used for the 4-stage non-inverting and inverting amplifiers.}
\label{tab:amplifiers}
\end{table}

\begin{figure}[ht]
\centering
\includegraphics[width=3.3in]{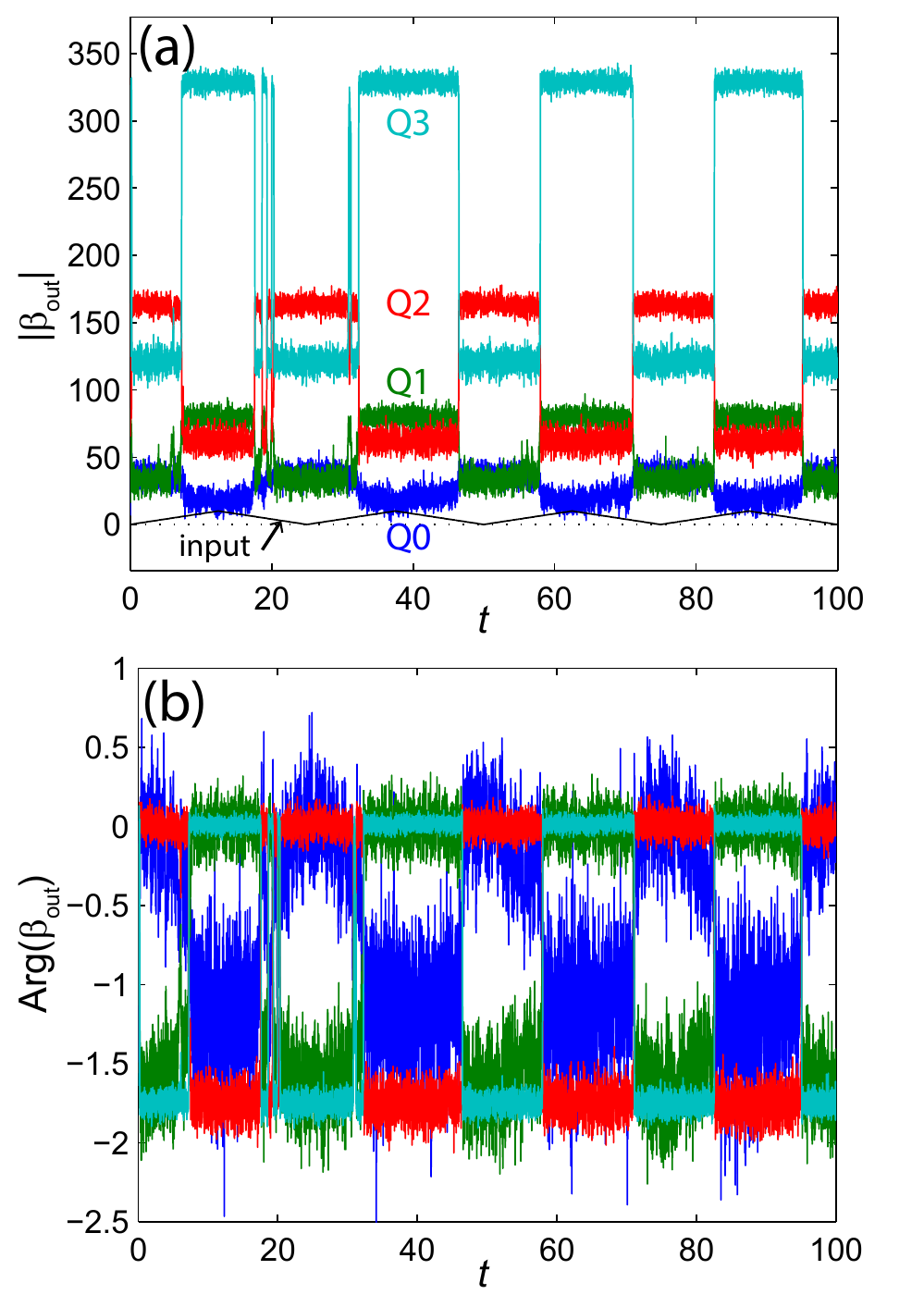}
\caption{(a) Simulated output amplitudes from stages 0, 1, 2, and 3 (labeled Q0, Q1, etc.) for four cascaded inverting amplifiers, using a triangle-wave input (black curve) with amplitude varying linearly between 0 and 10.  The fields are averaged over a time interval of 0.01.  (b) Corresponding output phases.}
\label{fig:amplifier_outputs}
\end{figure}

Because of its threshold-like behavior, this amplifier has some digital signal restoration capability.  Here, we perform simulations with a linearly varying triangle-wave input, with amplitude between 0 and 10, to examine whether the signal restoration, in the presence of quantum noise, is sufficient for cascading. For multiple stages, since each stage receives a larger input amplitude, the parameters for each amplifier stage must be chosen differently (Table~\ref{tab:amplifiers}) to match the switching thresholds with the expected low and high resonator inputs.  The simulated output field amplitudes of four cascaded inverting amplifiers are shown in Fig.~\ref{fig:amplifier_outputs}a.  We see that, despite the badly behaved input, the first stage (blue curve) has a clear switching behavior, although there is some variation in the output amplitude within the high and low states. In subsequent stages, the curves are quite flat in between switching events, and furthermore, the short-timescale noise due to quantum fluctuations does not appear to increase from one stage to the next.  The corresponding output phases are plotted in Fig.~\ref{fig:amplifier_outputs}b. The phase noise can be seen to decrease with each stage.

\begin{figure}[ht]
\centering
\includegraphics[width=3.3in]{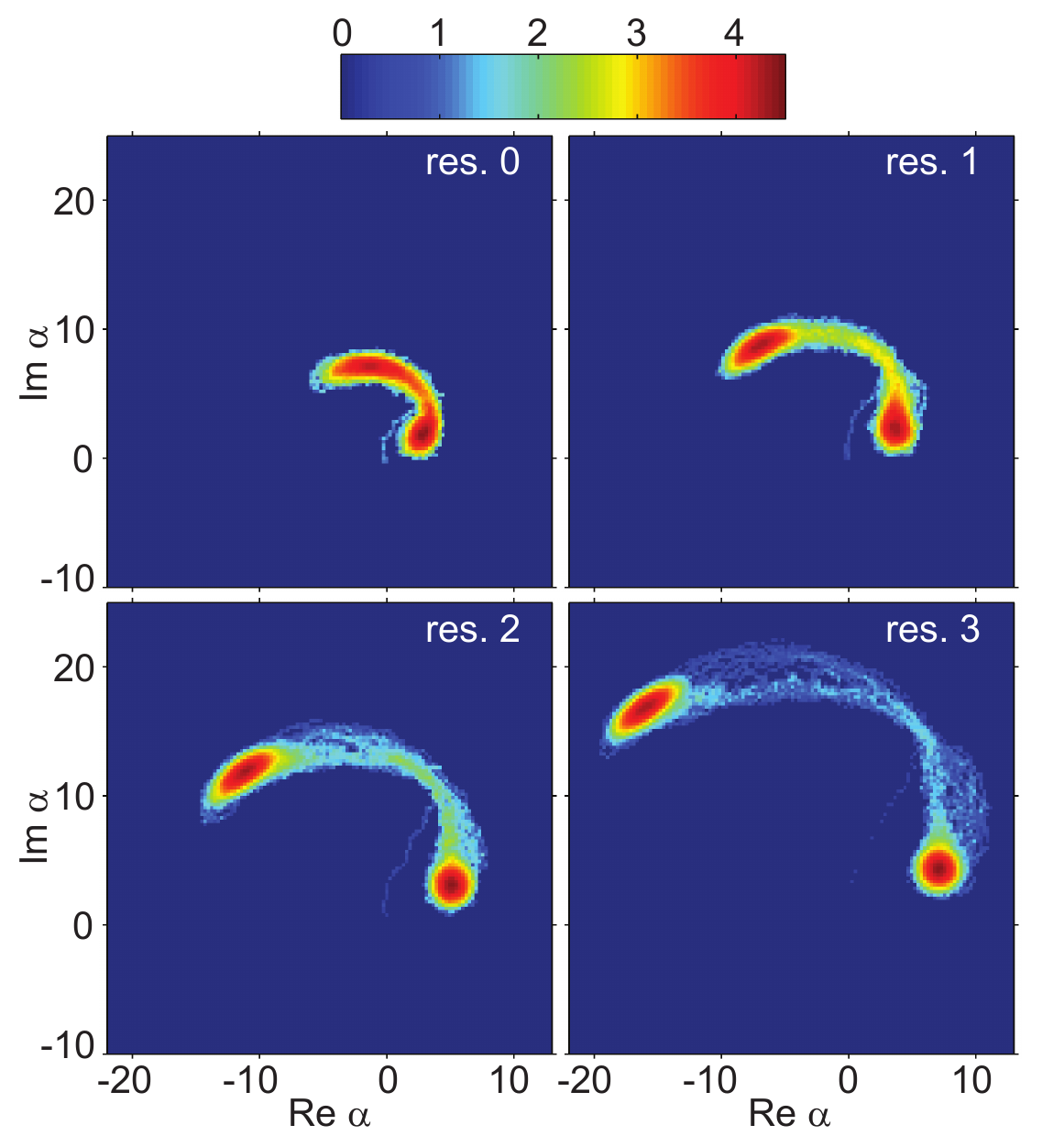}
\caption{Base-10 logarithm of the unnormalized, time-averaged distribution of the complex resonator fields in a four-stage inverting amplifier circuit, using a triangle-wave input.}
\label{fig:amplifier_wigner_inverting}
\end{figure}

\begin{figure}[ht]
\centering
\includegraphics[width=3.3in]{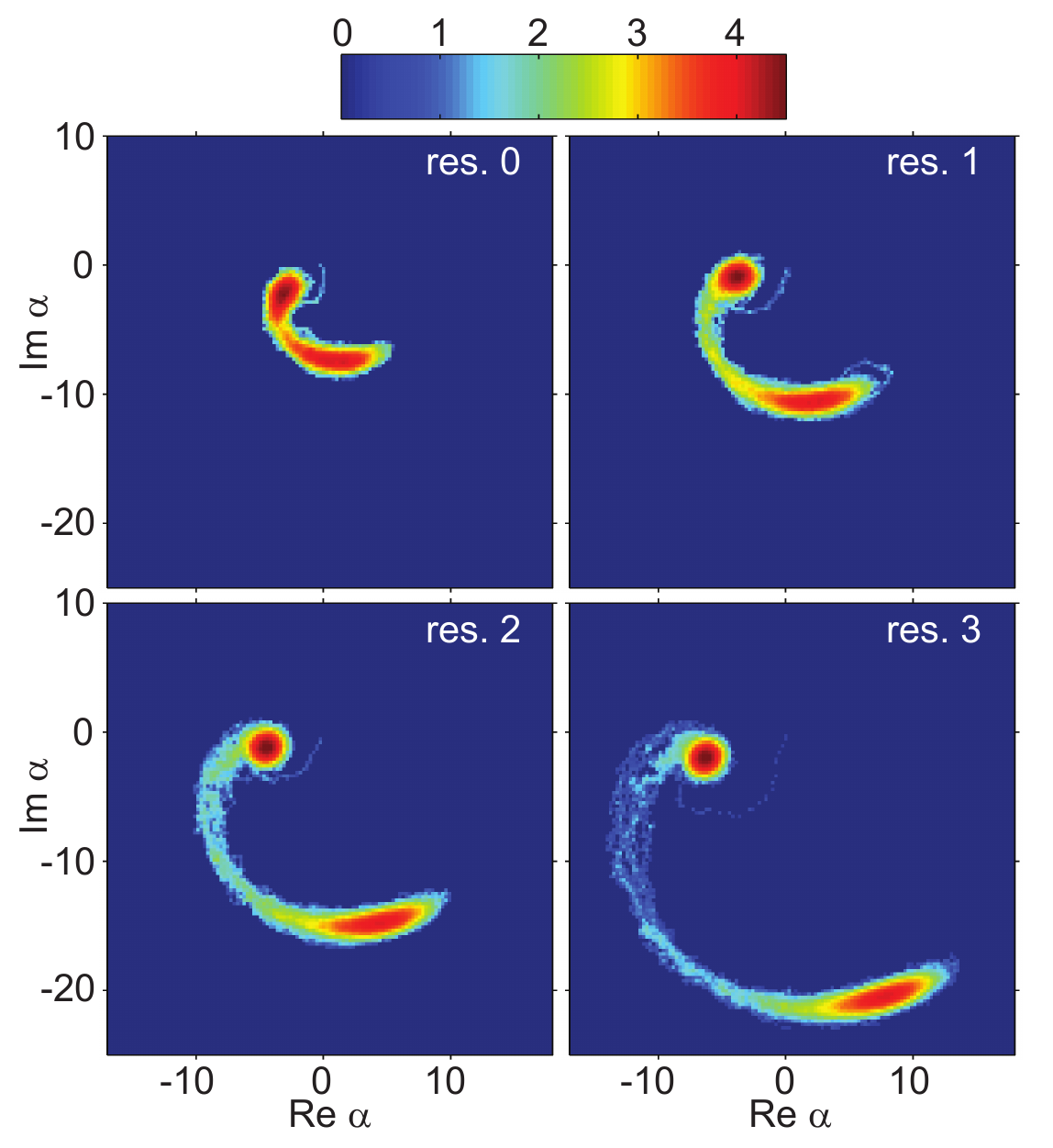}
\caption{Same as Fig.~\ref{fig:amplifier_wigner_inverting}, but using non-inverting amplifier stages.}
\label{fig:amplifier_wigner_noninverting}
\end{figure}

When performing these simulations, we noticed a markedly different behavior when we cascade inverting or non-inverting amplifiers.  This can be seen most easily in the time-averaged complex field distributions of the resonators for the two cases, which are plotted in Figs.~\ref{fig:amplifier_wigner_inverting} and~\ref{fig:amplifier_wigner_noninverting}. The figures use a logarithmic scale in order to cover $>4$ orders of magnitude, allowing the faint connections between the high and low states to be seen.  In either case, in the first amplifier stage we see that the high state has a larger phase variation than the low state, which is a general feature for Kerr resonators even for constant inputs.  In the inverting case, this noisier high state from the first stage leads to a low state in the second stage, which largely resets the noise.  In the non-inverting case, the high state from the first stage leads to another high state in the second stage, allowing the phase noise to propagate further, even though it is diluted by the contribution of the next auxiliary coherent-state input.  As a result, the phase noise in the last stage is larger for the non-inverting amplifiers, compared with the inverting amplifiers. If each amplifier is driven with a constant input, $\langle \beta_\mathrm{in} \rangle = 10$, after the fourth stage the phase noise amplitude of the non-inverting amplifier chain is $53\%$ higher than for the inverting amplifier chain. This is part of the motivation for using inverting amplifiers in the fan-out circuits presented below.

\subsection{AND gate}
\label{sec:AND}
\begin{figure}[ht]
\centering
\includegraphics[width=3.3in]{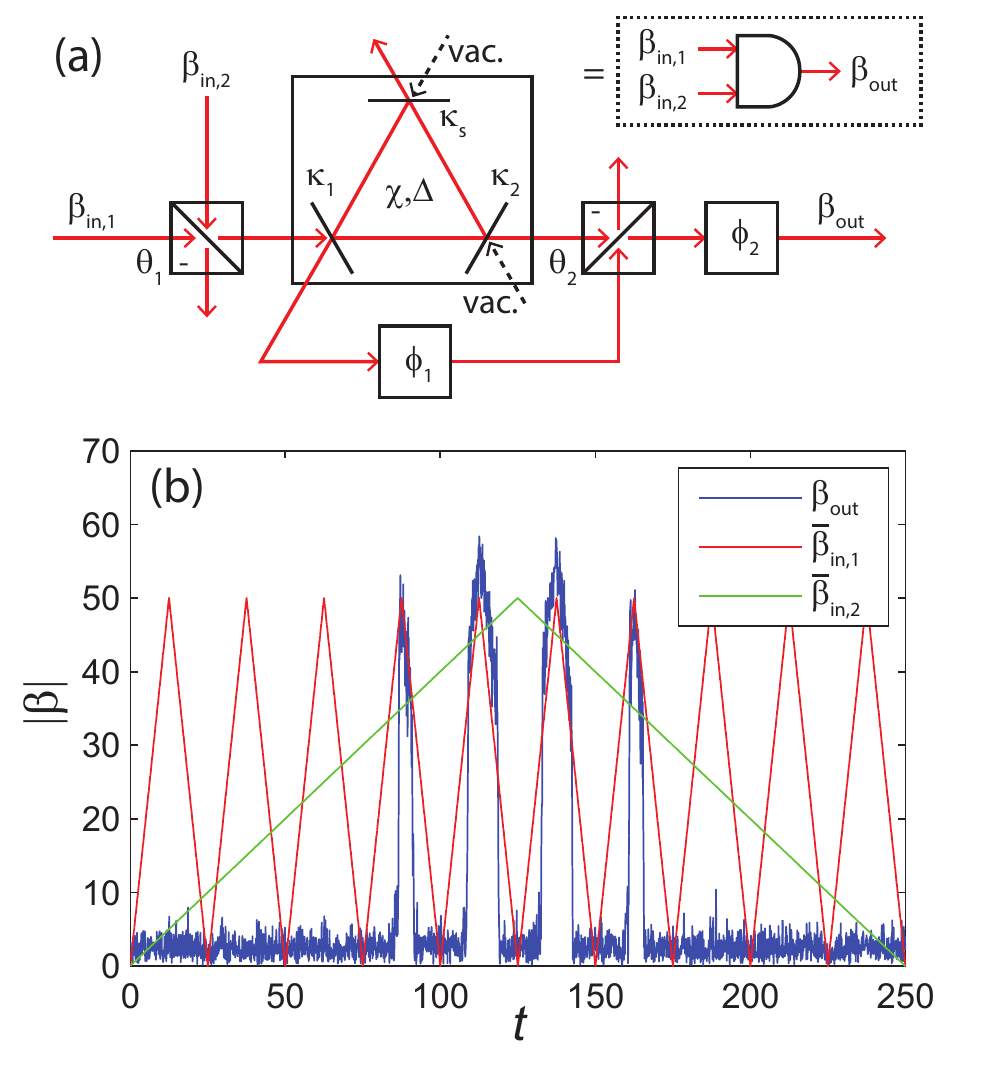}
\caption{(a) AND circuit based on a nonlinear Kerr resonator. (b) Simulated output field (blue) for triangle-wave inputs (red, green), with $E_\mathrm{high}=50$. The fields are averaged over a time inverval of 0.01.}
\label{fig:AND_circuit}
\end{figure}

The first logic gate we consider is the AND gate shown in Fig.~\ref{fig:AND_circuit}a.  This is the same circuit that was shown in Fig. 1a of Ref.~\cite{ref:mabuchi2011nia}, except that here we include an intrinsic resonator loss, which will be unavoidable in practical photonic integrated circuits.  Even if we have complete control over the resonator's output coupling rates, if we optimize for the lowest possible switching energy, then the intrinsic losses will be comparable to the coupling losses.  Suppose, for example, that we design a gate to work near the onset of bistability, with $|\Delta| = \kappa$.  The switching photon number will be close to the inflection point in the bistability curve, $n_\mathrm{switch} \approx | \Delta / (3\chi) |$. Using Eq.~\ref{eq:class_bistable} for the input field $\beta_\mathrm{in}$, and setting the minimum switching energy (in photon units) to be $U_\mathrm{switch} \approx |\beta_\mathrm{in}|^2/\kappa$, we find,
\begin{equation}
U_\mathrm{switch} \approx \frac{\kappa^2}{|\chi| \kappa_1}
\end{equation}
Setting $\kappa = 2\kappa_1 + \kappa_3$, where $\kappa_3$ is the intrinsic loss, and minimizing $U_\mathrm{switch}$ with respect to $\kappa_1$, we obtain $\kappa_3 = 2\kappa_1 = \kappa/2$, so that the intrinsic loss is half of the total loss. However, with such a large relative intrinsic loss, it is difficult to design circuits in which the output field amplitudes are as large as the input amplitudes. Thus, we have backed off from the optimum and used $\kappa_3 = 0.2 \kappa$.  The circuit parameters used to simulate the AND gate are given in Table~\ref{tab:gates}.

The AND gate works as follows: the two inputs interfere on a 50-50 beamsplitter.  Only if both inputs are high (and in phase), the beamsplitter output is large enough to exceed the switching threshold of the resonator.  We could simply use the output from resonator mirror 2 as the final output, as we did in the amplifiers discussed above. However, the performance can be improved by taking the output from resonator mirror 1, adjusting its phase, and interfering it with the output from resonator mirror 2 on a second beamsplitter. We adjust the phase $\phi_1$ and the mixing angle $\theta_2$ so that, when only one of the circuit inputs is ``high'', the signals entering the second beamsplitter interfere destructively, giving a ``low'' output close to zero.  The final phase shift $\phi_2$ is chosen so that the ``high'' output has its phase close to zero.  

Fig.~\ref{fig:AND_circuit}b shows the simulated operation of the AND gate.  Triangle waves are used to drive both inputs, to test the circuit's capability for digital signal restoration.  The ``high'' output amplitude slightly exceeds the designed input ``high'' amplitudes. A separate simulation, in which the circuit was driven by a square-wave signal, indicates propagation delays (for the output to cross a level halfway between the steady-state low and high levels) ranging from $\tau = 0.023 = 1.2 / \kappa$ for the fastest $\{1,1\}\rightarrow \{0,0\}$ input transition, to $\tau = 0.091 = 4.6/\kappa$ for the slowest $\{0,1\}\rightarrow \{1,1\}$ transition. Because of the direct path from the input to the output (through $\phi_1$), the output can also exhibit short spikes if the inputs transition suddenly from $\{0,0\}$ to $\{1,0\}$.

\begin{table}[h]
\centering
\begin{tabular}{|c|c|c|c|}
\hline
   & AND & Fanout & Latch \\
\hline
design $\beta_\mathrm{in}$ & $\{0, E_\mathrm{high}\}$ & $\{0, E_\mathrm{high}\}$ & $\{0, E_\mathrm{high}\}$ \\
$\beta_c$ &  & $-2.6 E_\mathrm{high}$ & $1.75 e^{-i\phi_1} E_\mathrm{high}$ \\
$\chi$ & $-653.4/E_\mathrm{high}^2$ & $-348.48/E_\mathrm{high}^2$ & $-512.5/E_\mathrm{high}^2$ \\
$\kappa_1$ & 20 & 20 & 20 \\
$\kappa_2$ & 20 & 20 & 20 \\
$\kappa_3$ & 10 & 10 & 10 \\
$\Delta$ & 50 & 50 & 50 \\
$t_1 = \cos \theta_1$ & 0.707 & 0.707 & 0.707 \\  
$t_2 = \cos \theta_2$ & 0.89  & 0.89  & 0.629 \\
$t_3 = \cos \theta_3$ &       & 0.707 & 0.829 \\
$\phi_1$ & -1.39 & -1.45 & 2.72 \\ 
$\phi_2$ &  2.65 & -0.46 & 0.14 \\
$\phi_3$ &       &       & 2.34 \\
\hline
\end{tabular}
\caption{Parameters used for the basic circuits. Here, $E_\mathrm{high}$ is the ``high'' level for external inputs. The simulations used $E_\mathrm{high} = 50$ (resonator photon number $\sim 100$) or $E_\mathrm{high} = 20$ (resonator photon number $\sim 20$).}
\label{tab:gates}
\end{table}

\subsection{Inverting fan-out}
\label{sec:fanout}
The AND gate described above has only a single optical output. Splitting this in two before sending it to other gates (such as more AND gates) would fail, since the amplitudes would be a factor of $1/\sqrt{2}$ smaller, and would be near or below the switching thresholds of the subsequent gates. Here we briefly introduce an inverting amplifier circuit that can be used both as a 2x fan-out, and as a NOT gate to complement the AND gate above, allowing for universal combinatorial logic.

\begin{figure}[ht]
\centering
\includegraphics[width=3.3in]{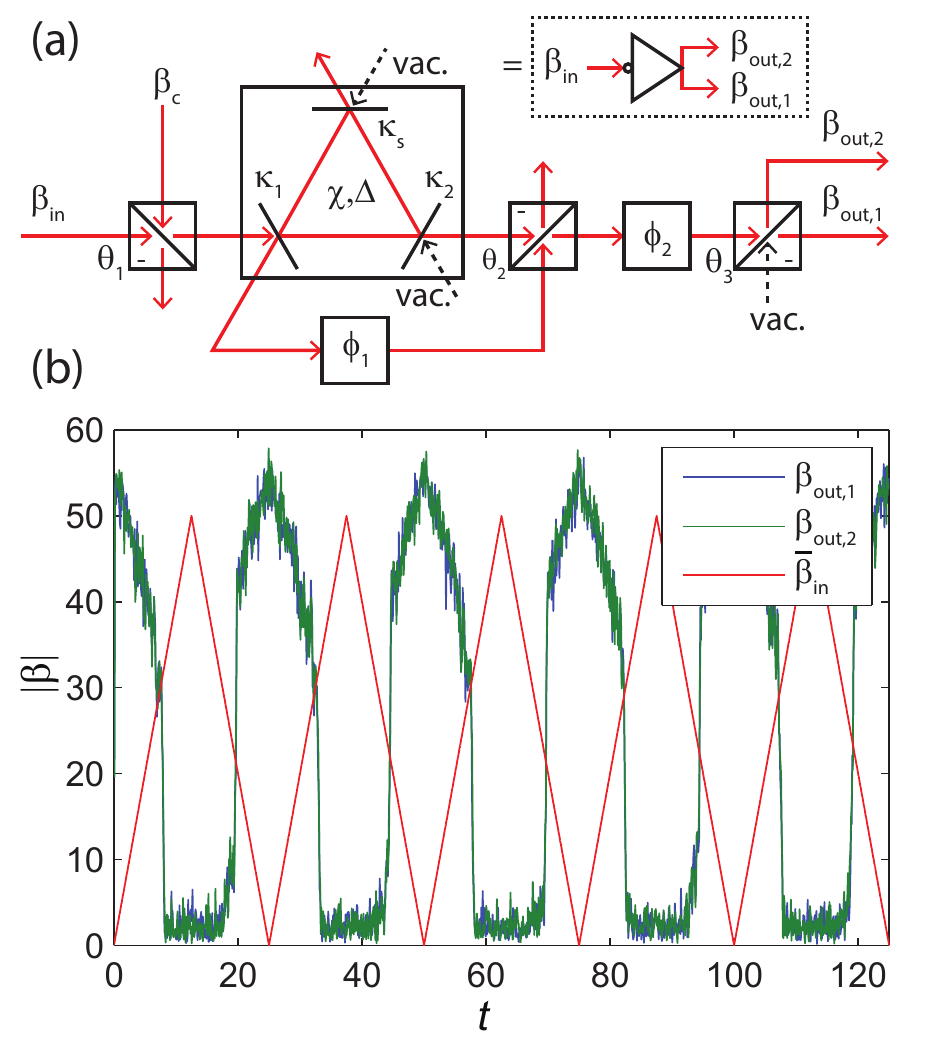}
\caption{(a) Inverting 2x fan-out circuit based on a nonlinear Kerr resonator. (b) Simulated output fields (blue, green) for a triangle-wave input (red), with $E_\mathrm{high}=50$. The fields are averaged over a time inverval of 0.01.}
\label{fig:NOT_circuit}
\end{figure}

Fig.~\ref{fig:NOT_circuit}a shows the inverting fan-out circuit, which is similar to the AND gate above, with a few differences. One of the inputs is replaced by a constant coherent drive which has a larger amplitude and is $180^\circ$ out of phase with respect to the remaining input. The resonator nonlinearity is also decreased in order to increase the switching threshold. In an integrated photonics implementation, the per-photon nonlinearity could be changed most easily by varying the resonator length.  The circuit parameters are given in Table~\ref{tab:gates}.  As a result of these changes, the output field in the high state has sufficient amplitude that it can be divided in two at a final 50-50 beamsplitter, yielding two outputs with amplitudes slightly above the designed high level.  Fig.~\ref{fig:NOT_circuit}b shows the simulated behavior of this circuit for a triangle-wave input, demonstrating the digital signal restoration capability of this circuit. For a square-wave input, the propagation delays (for mid-level crossing) are $\tau = 0.105 = 5.3/\kappa$ for the upward output transition and $\tau = 0.058 = 2.9/\kappa$ for the downward transition.

If fan-out to many outputs is required, one could design special circuits similar to the cascaded amplifiers described in~\ref{sec:amplifiers}.  However, for the present demonstration, for simplicity we shall limit the number of primitive components, cascading the 2x fan-out as needed to create additional copies of a signal.

\subsection{Latch}
\label{sec:latch}
\begin{figure}[ht]
\centering
\includegraphics[width=3.3in]{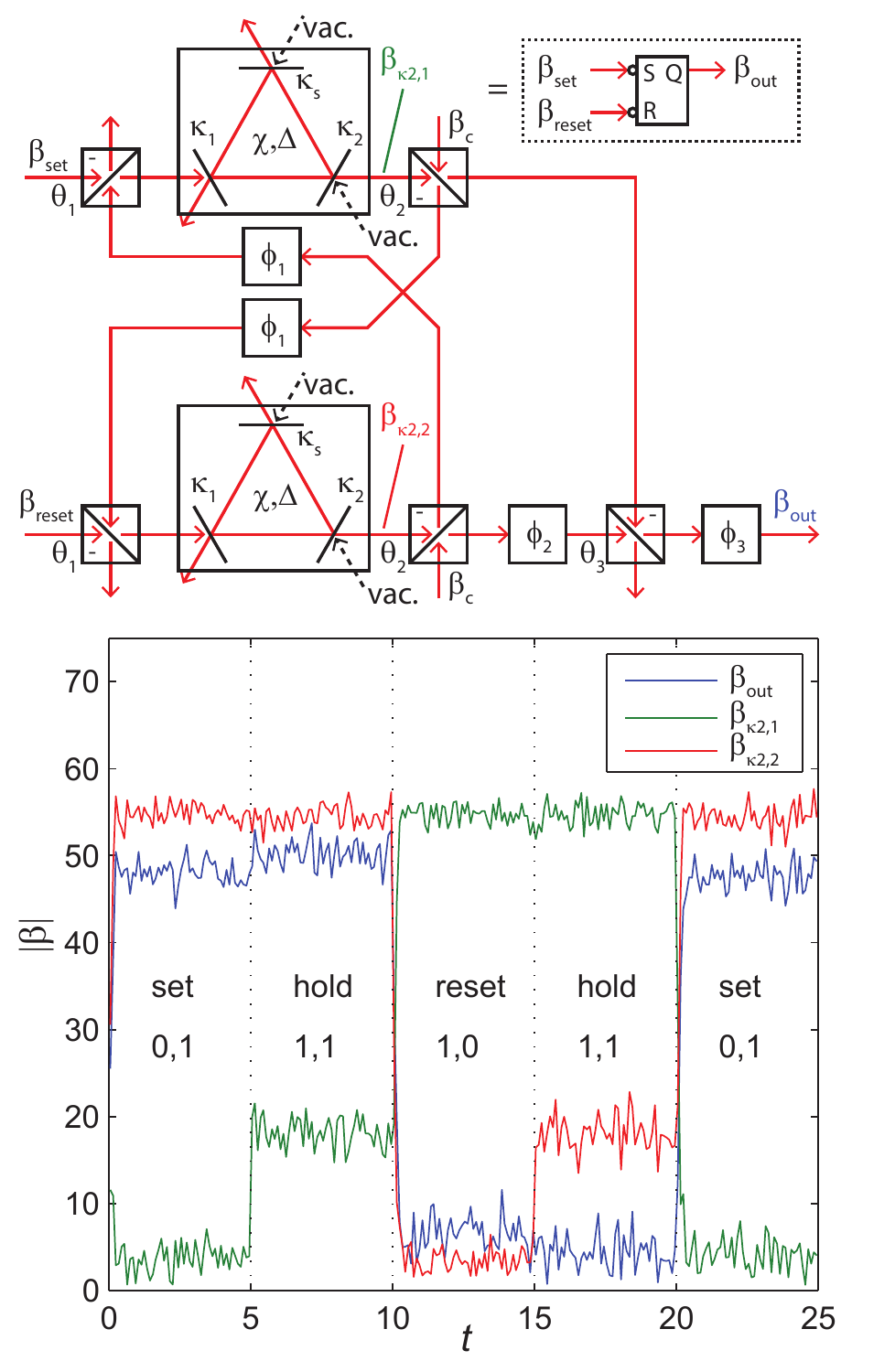}
\caption{(a) Latch circuit using two nonlinear Kerr resonators (see Ref.~\cite{ref:mabuchi2011nia}). (b) Simulated final output field (blue), and the fields exiting from the $\kappa_2$ ports of resonators 1 and 2 (the upper and lower resonators in the diagram, respectively; green and red curves).  For times in between the vertical lines, the two input levels are held constant at $\{0, 1\} E_\mathrm{high}$, as indicated, with $E_\mathrm{high}=50$. The fields are averaged over a time inverval of 0.01.}
\label{fig:latch_circuit}
\end{figure}

To add memory to our circuits, we start with the $\overline{SR}$-NAND latch from Ref.~\cite{ref:mabuchi2011nia}.  The main change we made to the version shown in Fig.~\ref{fig:latch_circuit}a was to incorporate intrinsic resonator losses, which required adjusting the other parameters (see Table~\ref{tab:gates}) so that the resonators operate closer to their switching thresholds. We might expect this to make the circuit more sensitive to noise.

The behavior of this circuit, simulated in Fig.~\ref{fig:latch_circuit}b, can be understood as follows. Let us name the upper and lower resonators in the diagram resonators 1 and 2, respectively. If $\beta_\mathrm{reset}$ is high and resonator 1 is off, the coherent input $\beta_c$ entering from the top of the diagram interferes constructively with $\beta_\mathrm{reset}$, so that the input to resonator 2 exceeds its switching threshold, keeping it in its ``on'' state.  The feedback phase is chosen such that the $\kappa_2$ output of resonator 2, feeding back to the input of resonator 1, interferes destructively with the other inputs, keeping resonator 1 in its ``off'' state, independent of whether $\beta_\mathrm{set}$ is low (the ``set'' condition) or $\beta_\mathrm{set}$ is high (the ``hold'' condition).  On the other hand, if $\beta_\mathrm{set}$ is high and $\beta_\mathrm{reset}$ is low (the ``reset'' condition), resonator 1 turns on and resonator 2 turns off.  If $\beta_\mathrm{reset}$ then goes high again (the ``hold'' condition), resonator 1 stays on and resonator 2 stays off. Thus, in the ``hold'' condition, the system retains its previous state.  The maximum propagation delay (for mid-level crossing) is $\tau = 0.11 = 5.5/\kappa$.

For $E_\mathrm{high}=50$, resonator 1 contains approximately 0.5 photons in its undriven (set) state, 17 photons in its lower hold state, and 150 photons in its driven (reset) and higher hold states.  The corresponding output field amplitudes are $\beta_{\kappa_2,1} \approx \sqrt{\kappa_2 n} \approx \{3, 18, 55\}$. The parameters of the final beamsplitter, which combines the outputs from the two resonators, and the associated phase $\phi_2$, are chosen so that the final output is as close as possible to $E_\mathrm{high}$ whenever resonator 2 is in its high state, and as close as possible to zero otherwise.

\begin{figure}[ht]
\centering
\includegraphics[width=3.3in]{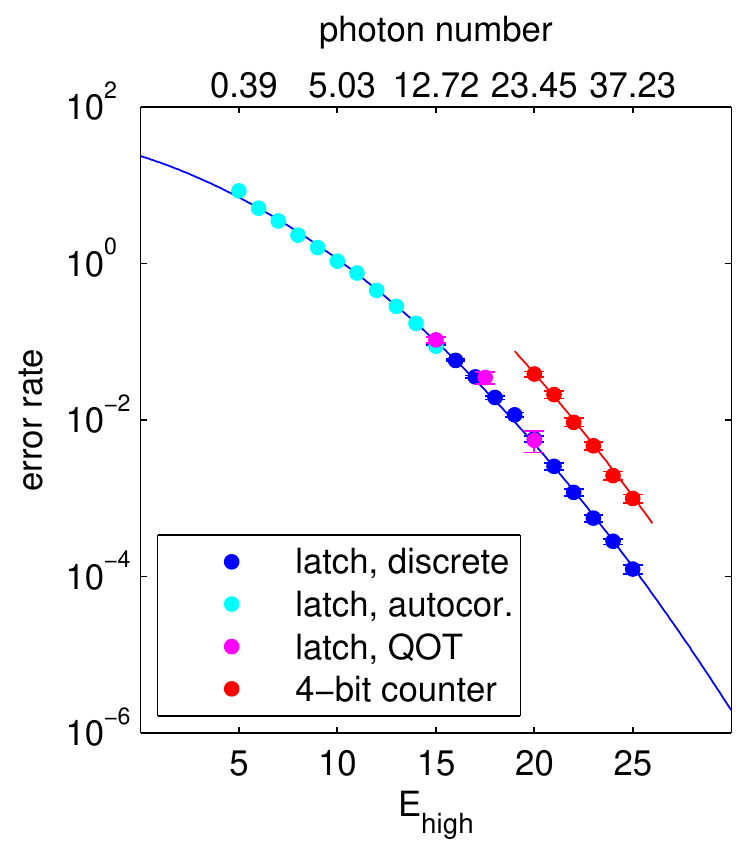}
\caption{Dark blue points: simulated spontaneous jump rates for a latch circuit in the ``hold'' condition, plotted as a function of the input field amplitude $E_\mathrm{high}$ (other circuit parameters are tied to $E_\mathrm{high}$ as described in Table~\ref{tab:gates}).  The error bars assume Poisson statistics for the number of detected jumps. Cyan points: simulated jump rate in the low-field regime, estimated by fitting an exponential decay to the resonator field autocorrelation.  Blue curve: quadratic fit to the logarithm of the jump rate.  Magenta points: results obtained using the quantum optics toolbox~\cite{ref:tan1999ctq,ref:mabuchi2011nia}.  Red points: simulated jump rate for the 4-bit counter circuit. Red curve: same as the blue curve but multiplied by 8 (see text). The top x-axis indicates the mean photon number, corresponding to $E_\mathrm{high}$, for a resonator in its ``on'' state during the ``set'' condition.}
\label{fig:latch_errors}
\end{figure}

When $E_\mathrm{high} \le 25$, spontaneous jumps between the two bistable states of the ``hold'' condition occur with sufficient frequency to allow accurate estimation of the jump rate, as shown in Fig.~\ref{fig:latch_errors}. When $E_\mathrm{high} \le 15$, this rate becomes too large to allow accurate state determination in between jump events, but we have extended the estimate to lower fields by fitting an exponential decay to the resonator field autocorrelation function. We should point out that, as $E_\mathrm{high} \rightarrow 0$ the semiclassical approximation is expected to become less and less accurate.

In solid-state implementations, our time units would likely correspond to intervals ranging from picoseconds to nanoseconds, and thus an acceptable error rate for computing could be estimated as $< 10^{-18}$. By fitting a quadratic polynomial to the logarithm of the jump rate, and extrapolating to higher input fields, we estimate that $E_\mathrm{high} \approx 54$ would be required to achieve this, corresponding to $\sim 177$ photons contained within a resonator in its ``on'' state during the ``set'' condition (this corresponds to $35\,\mathrm{aJ}$ at $\lambda = 1\,\mu\mathrm{m}$).

\subsection{Type-D flip-flop}
\label{sec:Dflipflop}
\begin{figure}[ht]
\centering
\includegraphics[width=3.3in]{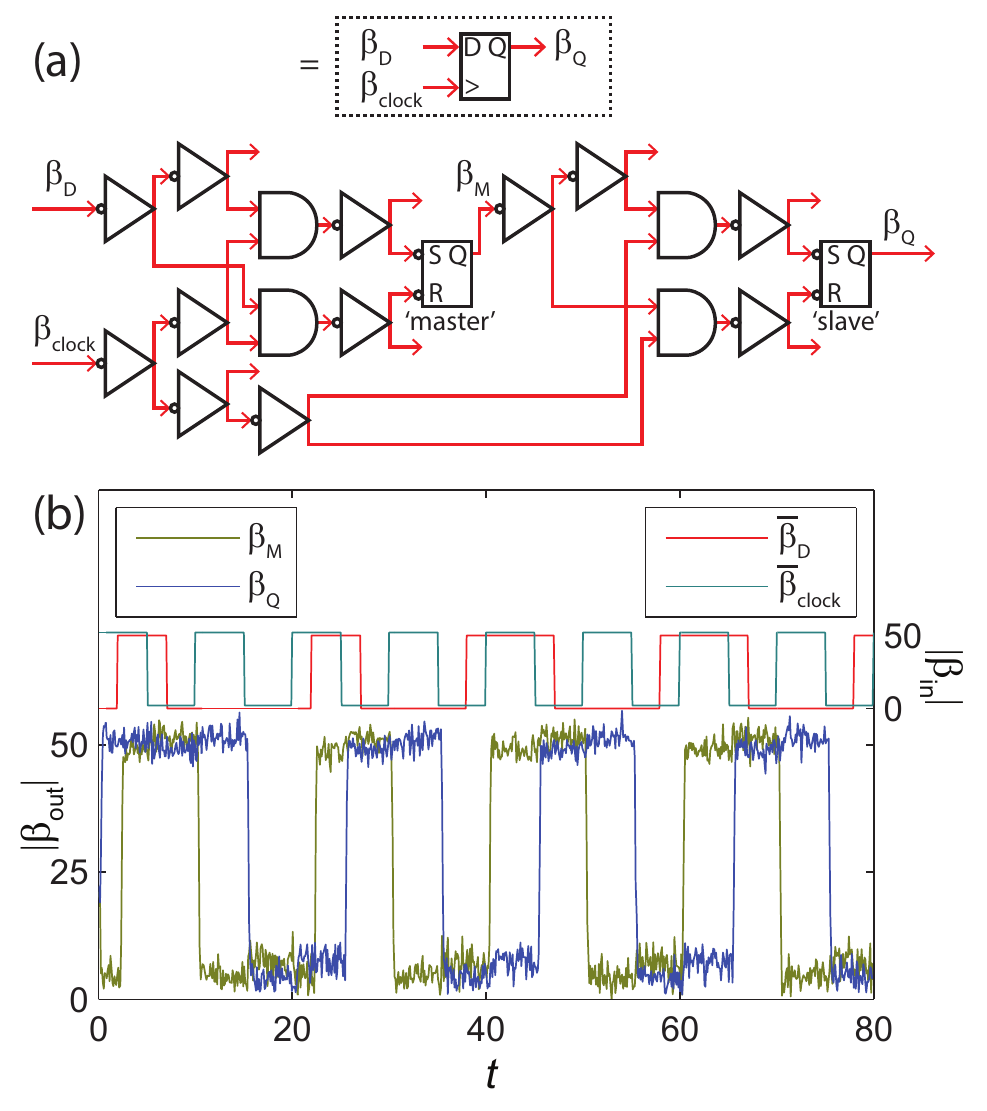}
\caption{(a) Schematic for a D flip-flop, built from the primitive circuits described above. (b) Simulated ``master'' output (yellow) and final ``slave'' output (dark blue) in response to the main input (red) and clock (light blue) square-wave signals. The simulation used $E_\mathrm{high}=50$, and the plotted fields are averaged over a time interval of 0.01.}
\label{fig:flipflop_circuit}
\end{figure}

We next combine the primitive components defined above to make a clocked memory component that functions the same as the $D$ flip-flop in electronics~\cite{ref:horowitz1989ae}.  The circuit schematic is shown in Fig.~\ref{fig:flipflop_circuit}a (see the previous figures for gate symbol definitions). This circuit is built around two $\overline{SR}$-NAND latches, denoted as ``master'' and ``slave''. The main input is denoted $\beta_D$, the clock is $\beta_\mathrm{clock}$, the intermediate ``master'' output is $\beta_M$, and the final ``slave'' output is $\beta_Q$. AND gates, with the clock or its compliment as one of the inputs, are used to control when the latches can change states. When the clock is high, $\beta_D$ controls the state of the ``master'' latch, while the ``slave'' latch is frozen.  When the clock goes low, the ``master'' latch is frozen, but its state is transferred to the ``slave'' latch.  Numerous inverting fan-out gates are also required, either to divide or invert the various signals. Since these components account for more than half of the resonators in the circuit, replacing them with modified AND gates or improved amplifiers would be one of the more straightforward ways to optimize this circuit.

The simulated circuit dynamics are shown in Fig.~\ref{fig:flipflop_circuit}b. For the first two input pulses (red), the rising edges occur when the clock (light blue) is already high, and thus the master latch (yellow) switches at the input rising edge. For the second two input pulses, the rising edges occur before the clock is high, and thus the master latch switches at the clock rising edge. The slave latch (dark blue) always transitions at the falling clock edge.  Close examination shows that propagation delays (for mid-level crossing) of up to $\tau = 0.51 = 25.5/\kappa$ can occur following clock edges. This is approximately as one would expect from summing the individual component delays given above. For the slave transition, the path from the clock to the output passes through 4 fan-outs (with two upward and two downward transitions), an AND gate (upward), and a latch (can be upward). This circuit includes 20 resonators, 54 vacuum inputs, 16 non-vacuum coherent-state inputs, 54 beamsplitters, and 40 phase shifters. It is our first example of a circuit too large to simulate using known full-quantum methods.

\subsection{4-bit counter}
\label{sec:counter}
\begin{figure}[ht]
\centering
\includegraphics[width=2.0in]{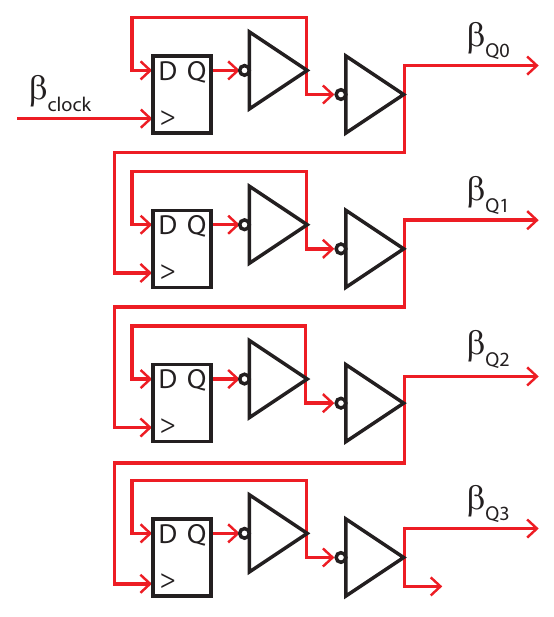}
\caption{Circuit diagram for a 4-bit ripple counter.}
\label{fig:counter_circuit}
\end{figure}

As a simple application of the D flip-flop, here we demonstrate a 4-bit ripple counter. The circuit, shown in Fig.~\ref{fig:counter_circuit}, contains 4 flip-flops, each representing one of the bits.  The output of each flip-flop is inverted and fed back into its main input, causing its output state to toggle at each falling clock edge. Additionally, the output of a given flip-flop serves as the clock signal of the flip-flop representing the next-higher-order bit. Thus, two inverting fan-out components are required at each stage to generate the required copies.

\begin{figure}[ht]
\centering
\includegraphics[width=3.3in]{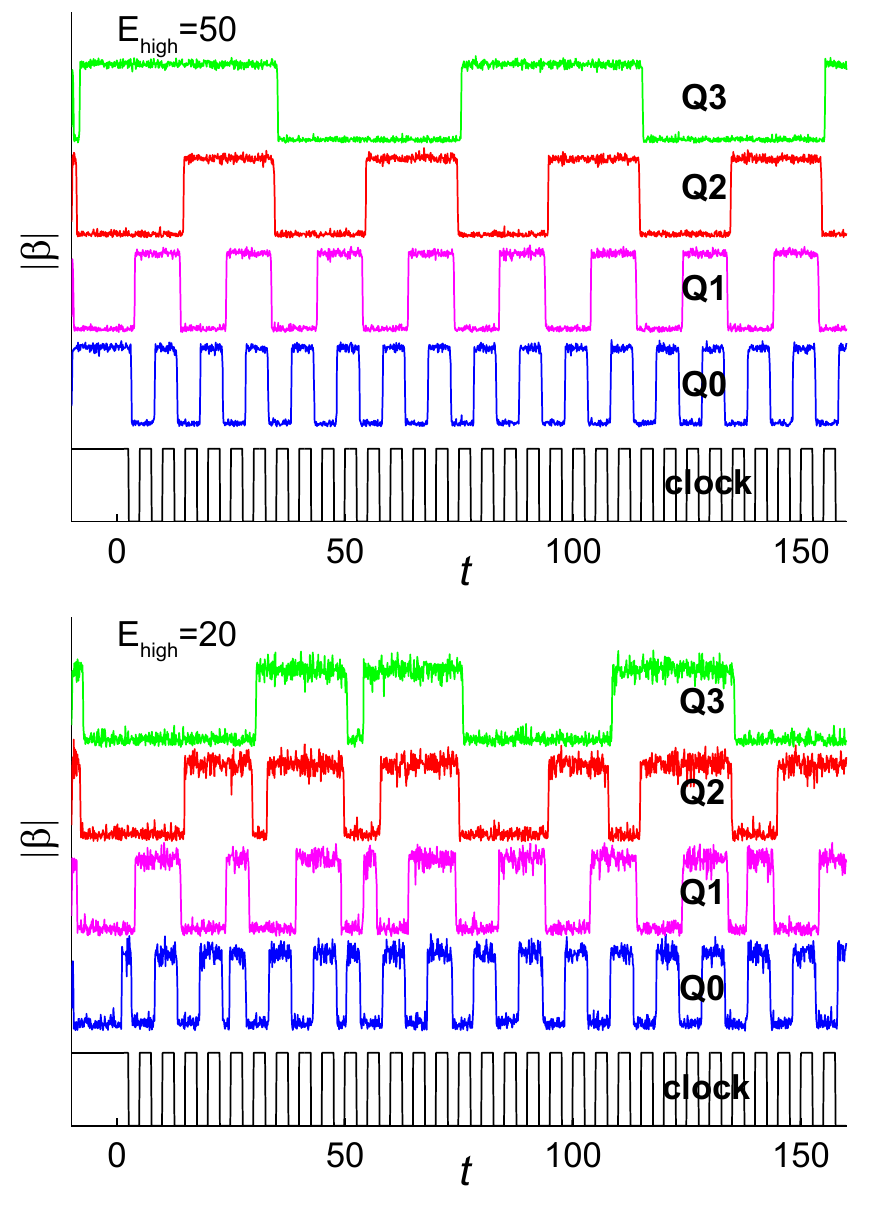}
\caption{Top: Simulated outputs of the 4-bit ripple counter for $E_\mathrm{high}=50$. Bottom: outputs for $E_\mathrm{high}=20$. The signals are averaged over a time interval of 0.1.}
\label{fig:counter_outputs}
\end{figure}

When flattened into its primitive components, this circuit contains 88 resonators, 240 beamsplitters and 176 phase shifters, and requires 233 vacuum inputs and 72 non-vacuum inputs.  The simulated behavior of the output fields is shown in Fig.~\ref{fig:counter_outputs}.  For $E_\mathrm{high}=50$, although the bit values start out with random values, the behavior for $t>0$ is exactly as expected.  However, for $E_\mathrm{high}=20$, random error events occur rather frequently. The estimated error rates for several values of $E_\mathrm{high}$ are plotted in Fig.~\ref{fig:latch_errors}. Since the counter circuit contains 8 latches, we expect the error rate to be at least 8 times larger than the error rate for a single latch. The good agreement between the counter error rates and the red curve in Fig.~\ref{fig:counter_outputs} (which is 8 times the fitted latch error rate) suggests that spontaneous jumps in the latch circuits under the ``hold'' condition are the most important error source in the counter circuit.

\begin{figure}[ht]
\centering
\includegraphics[width=3.3in]{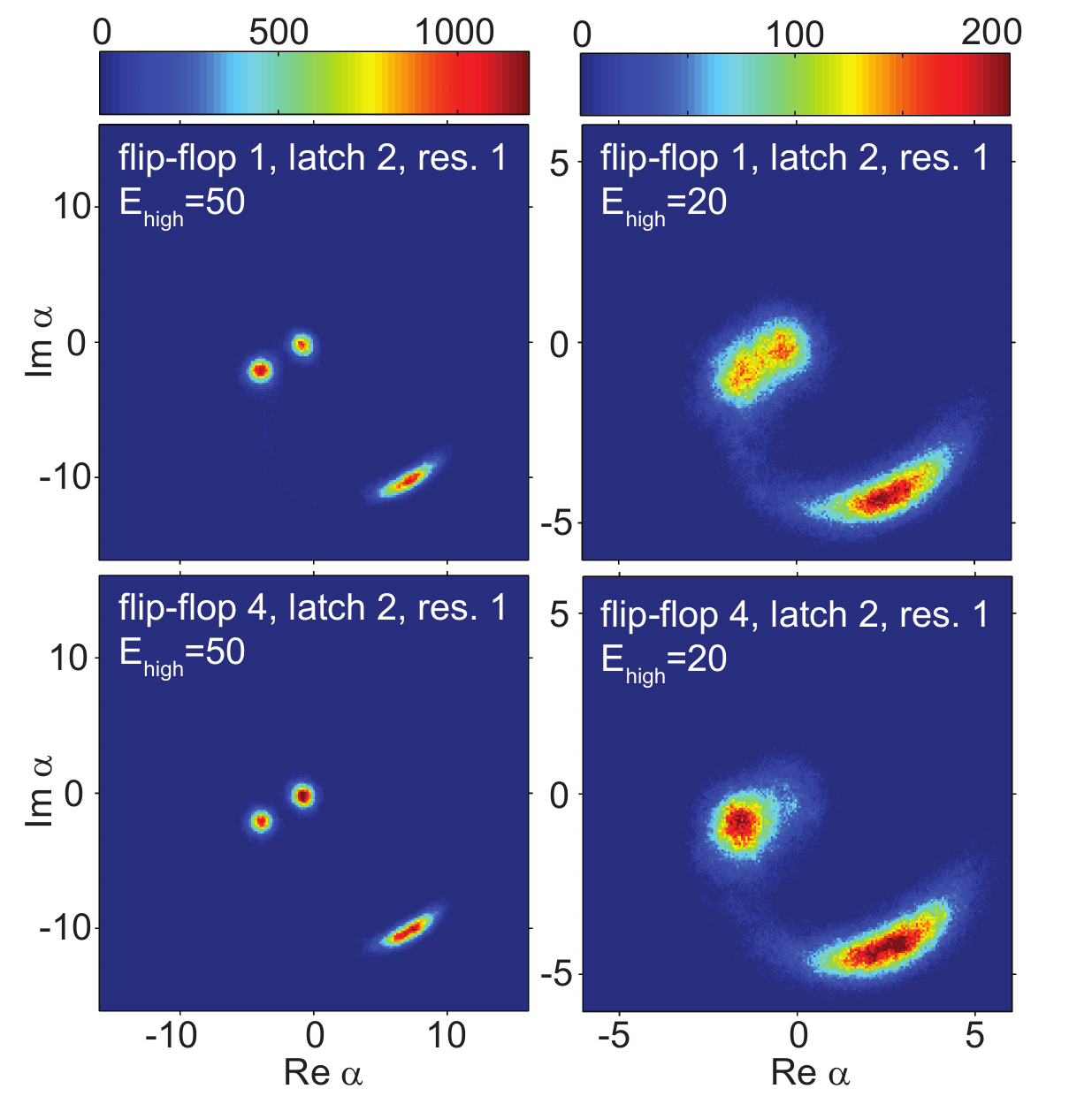}
\caption{For the 4-bit ripple counter, the time-averaged Wigner functions of the first resonator in the second latch of flip-flop 1 (upper-left) and flip-flop 4 (lower-left) for $E_\mathrm{high}=50$; and the same for $E_\mathrm{high}=20$ (upper-right and lower-right).}
\label{fig:counter_wfuns}
\end{figure}

It is also interesting to look at the Wigner functions of the resonators to see if the quantum noise grows as signals propagate through such a large circuit.  The time-averaged Wigner functions for two selected resonators are shown in Fig.~\ref{fig:counter_wfuns}.  One of the resonators is in the first flip-flop, and the other is in the last flip-flop. The small-scale quantum noise does not appear to grow. This is not surprising, considering the digital restoration properties demonstrated above for the basic gates.  Of course, large errors associated with quantum jumps between bistable states of a gate will propagate through the system, and we expect the overall error rate to scale linearly with the number of components.

\section{Outlook}

We have demonstrated a simulation approach suitable for studying quantum noise in large-scale, nonlinear photonic circuits. We used this model to simulate a digital counter incorporating gates and latches based on previous designs.  The results obtained so far suggest that the errors in the large-scale circuit are dominated by spontaneous jumps in the individual latches. From the error rates we extrapolate that in the current design, the resonators within the latch circuit must contain $\sim 180$ photons in their ``on'' state to achieve acceptable error rates for computing.

The circuits shown above were built from a very limited set of existing designs, chosen mainly to demonstrate the capabilities of this simulation approach. We are currently optimizing the circuits to make them more experimentally realizable. Initial results indicate that the number of resonators can be greatly reduced, by factors as large as 5 in some cases.  At the same time, we are working to make the circuits sufficiently tunable (by adjusting the amplitudes and phases of external inputs) to accommodate random variations in phases, resonant frequencies, and coupling strengths associated with fabrication imperfections.

At the same time, we are working to increase the capabilities of our model. We have recently incorporated the semiclassical equations used here into the QHDL software framework~\cite{ref:tezak2012spc}, which will allow a single tool to perform both semiclassical and full-quantum simulations for a given circuit, and also allows for graphical construction of circuits. We are also working to expand the model to include quantum noise associated with other kinds of nonlinearities. Carrier-based nonlinearities are of particular interest for experiments. While the Kerr model is the simplest, requiring a single degree of freedom per resonator in our circuits, we have found that the general design principles we applied to constructing the logic gates based on a Kerr nonlinearity fully translate to any sufficiently strong optical non-linearity, be it absorptive or dispersive.

\section{Acknowledgements}
This work was supported by the Defense Advanced Research Projects Agency under Agreement No. N66001-12-2-4007.


\end{document}